\documentclass{article}

\usepackage[T1]{fontenc}
\usepackage[utf8]{inputenc}
\usepackage{lmodern}
\usepackage{microtype}

\usepackage[margin=1in]{geometry}  
\usepackage{amsmath,amssymb,amsfonts,amsthm,latexsym,mathrsfs,bbm}
\usepackage{graphicx}
\usepackage{caption}
\usepackage{subcaption}
\usepackage{wrapfig}
\usepackage{longtable}
\usepackage{booktabs}
\usepackage{multirow}
\usepackage{array}
\usepackage{float}
\usepackage{pdflscape}             
\usepackage{enumitem}
\usepackage{setspace}
\usepackage{framed}
\usepackage{comment}
\usepackage{authblk}
\usepackage{url}
\usepackage[ruled,vlined,linesnumbered]{algorithm2e}

\usepackage{natbib}

\usepackage[unicode=true,colorlinks=true,linkcolor=blue,citecolor=blue,urlcolor=blue]{hyperref}

\begin{document}
\title{Modelling peaks over thresholds in panel data: a two-level grouped panel generalized Pareto regression}
\date{}
\author{Zefan Liu$^{a}$\textsuperscript{,}\thanks{Corresponding author. E-mail address: \texttt{zefan.liu@stat.ubc.ca}} and Natalia Nolde$^{a}$  \\$^{a}$ Department of Statistics, University of British Columbia, Canada}

\maketitle
\begin{abstract}
Panel data arise in a wide range of application areas, and developing modelling methods for extreme values under such a setup is essential for reliable risk assessment and management. When choosing to model the marginal distributions of univariate extremes, one may wish to balance the flexibility in capturing the heterogeneity among margins and the efficiency of estimation. This can be achieved through a combination of regression techniques and assuming a latent group structure based on parameter values, which needs to be estimated from data. Building on an existing method, we propose a two-level grouped panel generalized Pareto regression framework, which models peaks over high thresholds in panel data. While retaining the wide applicability of the original modelling strategy, which is largely domain-knowledge-free, our new methodology uses the information of extreme events more exhaustively and allows the exploration of a broader model space, where parsimony and good model fit can be achieved simultaneously. We also address several estimation challenges associated with high-dimensional optimization and group structure identification. The finite-sample performance of our methodology is carefully evaluated through simulation studies. With an application to the summer river flow data from 31 stations in the upper Danube basin, we show that our methodology can effectively improve estimation efficiency while discovering patterns in the tail behavior that can be omitted by domain-knowledge-based regionalization and the existing method.
\end{abstract}

\newpage
\setcounter{page}{1}

%=========INTRODUCTION=========
\section{Introduction}
\label{Introduction}

For modeling the behavior of extreme events from a probabilistic perspective, extreme value theory (EVT) provides statisticians with a theoretical foundation. For example, one of the most foundational results in univariate EVT states that, under certain conditions, the limiting distribution of the maximum of a sequence of independent and identically distributed (i.i.d.) random variables, as well as that of the conditional excesses over a high threshold, must belong to specific parametric families. Building on these classic results and their extensions, various practical methods have been developed to analyze patterns of extreme events in nature, making EVT invaluable in applications across environmental science, economics, finance, and many other fields.

In many scenarios of extreme value study, repeated observations for the same subjects or at the same locations over time are available, resulting in the formation of panel data. When the marginal distributions of extremes in panel data are studied, it is important to properly account for the potential heterogeneity across margins. In addition to incorporating covariates through regression techniques (see, e.g., \cite{spatial_extremes_modelling} and \cite{Youngman_2019}), a group structure across subjects—determined by the similarity in parameters—may also be assumed. Besides allowing additional model flexibility, compared to fitting a separate model for each subject or location, such a modelling strategy also facilitates information pooling across subjects or locations, which helps to address the scarcity of extreme values. Focusing on univariate extremes, \cite{Dupuis_2023} study the case where a panel of block maxima is directly available. A grouped panel generalized extreme value (GEV) regression model is proposed, where the marginal distributions of block maxima are assumed to be well approximated by the GEV distributions; instead of specifying a group structure based on the prior knowledge on the associations among subjects, such as their geographical closeness, hydrological connection or climatological similarity (\cite{asadi2015extremes}, \cite{alila1999hierarchical}), \cite{Dupuis_2023} assume that the group label of each subject uniquely characterizes the regression coefficients associated with the marginal GEV distributions for this subject. A two-step maximization algorithm, which is of the expectation-maximization (EM) algorithm type, is proposed to jointly estimate the group assignments of subjects and the corresponding regression parameters. The assumed cross-sectional dependence among margins is then accounted for by a sandwich covariance estimator, as is commonly done under the composite likelihood framework (see \cite{Overview_Composite_Likelihood} for a comprehensive review). The true number of latent groups is determined by comparing the Bayesian information criterion (BIC) of the identified group structures across a sequence of candidate number of groups. 

Motivated by advantages and wide applicability of the modelling idea of \cite{Dupuis_2023}, in this paper we present several extensions of the grouped panel GEV regression model. Specifically, we first consider its counterpart under the peaks-over-threshold framework, referred to as the grouped panel generalized Pareto (GP) regression model. Compared with the GEV version, this model is expected to utilize the information from extreme events more exhaustively (\cite{Coles2001Introduction}), especially when the events of interest are concentrated within a specific time interval of a long-duration period. While adopting an estimator similar to that of \cite{Dupuis_2023}, we provide additional details to support its validity, and a dependence-window-based sandwich covariance estimator is formulated. One limitation of the group structure assumption of \cite{Dupuis_2023} is that the estimated groups are determined by the joint similarity of regression coefficients associated with different distributional parameters of the GEV model. A similar assumption is made by \cite{de2023similarity}, in which time series are clustered based on the joint similarity of the scedasis functions and tail behavior of extreme values in these series. In some applications, this assumption may lack sufficient flexibility, which can then lead to a suboptimal marginal fit, the difficulty in interpreting the estimated group structure, and the omission of subtle but meaningful patterns in the data. We thus introduce a two-level grouped panel GP regression modelling framework, where separate group structures for the regression coefficients associated with the scale and shape parameters of the marginal GP models are allowed. This modification enables the exploration of a broader model space, where greater parsimony and better fit may be achieved simultaneously. When the shape parameters are treated as covariate-independent, a group structure that solely reflects the similarity across the tail behavior of subjects can be examined in isolation. We also offer solutions to handle the challenges in estimation associated with the two-level grouped model assumption. Specifically, given the potential numerical difficulty of parameter estimation due to high-dimensional optimization, we propose to use a block coordinate ascent (BCA) method (\cite{tseng2001convergence}) to compute the target estimates. This strategy is also incorporated in a multi-step maximization algorithm proposed for the joint estimation of the group structure and regression parameters (under a fixed group dimension pair). However, like many other EM-type algorithms, this algorithm is prone to converging to locally optimal solutions. When identifying the optimal group dimension pair by comparing the BIC values of candidate models, computational concerns may arise if a large number of algorithm runs is required for an exhaustive exploration of the model space. In addition, the identification performance on the group structure associated with the shape parameters is not overall satisfactory. We propose a two-stage approach in which the group structure for the shape parameters is estimated using an intuitive hierarchical clustering procedure. Compared to the original workflow proposed by \cite{Dupuis_2023}, this approach is expected to be overall less computationally intensive but more effective in capturing the group structure in the tail behavior of subjects. The proposed estimation methodology is carefully evaluated through a series of simulation studies, and satisfactory results are observed.

Under the overarching principle of striving for the best possible balance between estimation efficiency and model fit, we expect the two-level grouped panel GP regression model to be well-suited for a wide range of real-world applications where the behavior of extremes in panel data is of interest. For illustration, we apply our methodology to the daily river flow data of 31 hydrological stations within the upper Danube basin over summer months from 1960 to 2009. This dataset can be found in the supplementary material of \cite{asadi2015extremes} and has been used in several recent studies including \cite{hentschel2024statistical}, \cite{Dupuis_2023}, \cite{engelke2020graphical}, and \cite{asadi2015extremes}. Compared to the composite of all local GP models with no covariates included and the optimal grouped panel GP regression model, the two-level grouped model identified using the two-stage approach with hierarchical clustering is able to achieve a notably lower BIC, which is very close to that of the optimal model identified by repeatedly applying the multi-step maximization algorithm and then comparing the BIC values. While yielding distributional parameter and return level estimates that closely align with those from the local GP models across the majority of stations, this model is able to achieve a substantial reduction in the corresponding standard errors. We also carefully examine the group structures associated with the scale-related and shape-related parameters under this model, with the conclusions from \cite{asadi2015extremes} serving as references and a grouped panel GEV regression model adapted from the modeling choices of \cite{Dupuis_2023} used for comparison. Agreeing with \cite{asadi2015extremes}, the results based on data-driven regionalization methods also suggest that stations from mountain regions tend to have heavier tails. However, compared with the results under the grouped panel GEV regression model and prior-knowledge-based regionalization, our methodology is able to identify more subtle patterns in the tail behavior behind the data.

The remainder of the paper is organized as follows: In Section~\ref{Preliminaries} we review key results from (univariate) EVT that are used in the sequel. Section~\ref{methodology} outlines the model assumptions we make for the grouped panel GP regression model and the two-level grouped version. The suggested estimation methods for regression parameters, the associated standard errors, the latent group structure(s), and the optimal number of groups are then provided. In Section~\ref{Simulations} we present the results of the series of simulation studies that are designed to evaluate the finite-sample performance of our methods, and Section~\ref{Application} provides the details of the application. A summary of our work and directions for future research are outlined in Section~\ref{Conclusion}. Some relevant technical details and additional results from simulation studies and the application are provided in the supplementary material. R code to generate numerical results is available at \url{https://github.com/SteveLiu1723/Modelling-Peaks-over-Thresholds-in-Panel-Data}.

%=========PRELIMINARIES=========
\section{Preliminaries}
\label{Preliminaries}

In this section, we review key results from EVT on modelling excesses over a high threshold. The statistical procedure based on these results is known as the peaks-over-threshold (POT) approach. Consider a continuous random variable $Y$ with distribution function $F$. For a given threshold $u\in\mathbb{R}$, the conditional excess distribution function of $Y$ above $u$ is 
\begin{equation}
    F_u(z) := \mathbb{P}(Y-u\leq z | Y>u) 
    = \frac{F(z+u)-F(u)}{1-F(u)}. \label{eqt:cond_excess_dist_func}
\end{equation}
For a sufficiently high threshold $u$, $F_u$ can often be well modelled using a generalized Pareto (GP) distribution $GP(\sigma, \xi)$ with scale parameter $\sigma>0$ and shape parameter $\xi\in\mathbb{R}$, whose distribution function is given by
\begin{equation}
    H(z;\sigma,\xi) = 1- \left[1+\frac{\xi z}{\sigma}\right]^{-1/\xi}_+.
    \label{eqt:GP_cdf}
\end{equation}
This approximation is justified by the Pickands–Balkema–de Haan Theorem (\cite{Pickands_1975}, \cite{Dehhan_Balkema_1974}), which states that under suitable conditions, the following result holds:
\begin{equation}
    \lim_{u\rightarrow y_F}\sup_{0\leq z<y_F-u}\big|F_u(z)-H(z;\sigma(u),\xi)\big|=0,
    \label{eqt:Pickands–Balkema–De Haan Theorem}
\end{equation}
where $\sigma(u)$ is a positive function and $y_F$ is the upper end point of $F$. The parameter $\xi$ determines the tail behavior of $F$. Specifically, when $\xi<0$, $F$ has a finite upper end point; when $\xi>0$, $F$ is said to have a heavy tail, where the tail of $F$ becomes heavier as $\xi$ increases; when $\xi=0$, $F$ is said to have a light tail. 

For parameter estimation under the POT framework, we mainly focus on the likelihood-based method here (see \cite{hosking1987parameter} for a comprehensive review).  Let $\{Z_1,\ldots,Z_{T_u}\}$ be $T_u$ i.i.d. random excesses over threshold $u$, whose realizations are $\{z_1,\ldots,z_{T_u}\}$. Assuming that the distribution of threshold excesses can be well approximated by a GP distribution, the log-likelihood function is given by
\begin{equation}
    \ell_{GP}(\sigma, \xi;z_1,\ldots,z_{T_u}) = -T_u\log \sigma - \left(1+\frac{1}{\xi}\right)\sum^{T_u}_{i=1}\log \left(1+\frac{\xi z_i}{\sigma}\right),
    \label{eqt:GP_loglike}
\end{equation}
which requires $1+{\xi z_i}/{\sigma} > 0$ for all $i \in \{1,\ldots,T_u\}$. While the classical regularity conditions fail to hold over the global parameter space due to the parameter-dependent support of the GP distribution, the maximum likelihood estimators (MLE) are consistent and asymptotically normal under the constraint $\xi>-0.5$ (\cite{grimshaw1993computing}, \cite{Davison_Smith_1990}, \cite{hosking1987parameter}, \cite{Smith_1985}). This conclusion, along with similar ones under extended settings, is grounded in a foundational property of the GP distribution, which we introduce later. The case $\xi\leq-0.5$ is rarely encountered in practice, as it corresponds to a very short-tailed $F$. Analytical maximization of (\ref{eqt:GP_loglike}) is impossible, and numerical methods are required (\cite{Coles2001Introduction}).

In practice, it is often of interest to estimate a high quantile or return level for the random variable~$Y$. The $m$-observations return level, denoted by $y_m$, is defined as the level that is expected to be exceeded every $m$ observations. If excesses of $Y$ over a threshold~$u$ follow a GP distribution with parameters~$\xi$ and~$\sigma$, then
\begin{equation}
    \mathbb{P}(Y>y)=\mathbb{P}(Y>u)\left[1+\xi\left(\frac{y-u}{\sigma}\right)\right]^{-1/\xi}_{+}, \quad y>u.
    \label{eqt:prob_over_high_quantile}
\end{equation}
Setting $\mathbb{P}(Y>y)=1/m$ gives
\begin{equation}
    y_m=u+\frac{\sigma}{\xi}\Big[\big(m\mathbb{P}(Y>u)\big)^\xi-1\Big].
    \label{eqt:m-obs-returnlevel}
\end{equation}
In practical settings, return levels are often expressed on the annual scale for interpretability.

To obtain the MLE of $y_m$, all unknown parameters and quantities in (\ref{eqt:m-obs-returnlevel}) are replaced with their respective MLEs. The MLE of the exceedance probability $\mathbb{P}(Y > u)$ is simply the empirical proportion of sample points exceeding $u$. $\mathrm{Var}(\hat{y}_m)$ can be then estimated by the delta method. 

In the case when the underlying sequence of random variables is stationary, under a mild distributional mixing condition  (\cite{Leadbetter_1983_article}), the limiting result in the Pickands–Balkema–de Haan Theorem continues to hold. However, when excesses over a threshold exhibit clustering,  (\ref{eqt:GP_loglike}) is no longer the full log-likelihood. If one does not wish to model the dependence structure explicitly, then a common approach is to first decluster the excesses. However, declustering usually results in a loss of information. In addition, \cite{fawcett2007improved} show that the default choice of retaining only cluster maxima can lead to a severe bias in parameter estimates. As they suggest, an alternative approach is to use all excesses over a high threshold to estimate the parameters of the marginal GP distribution under a working assumption of independence, and then adjust the estimated standard errors to incorporate the dependence among the excesses. We will adopt this modelling choice in our work. 

%=========METHODOLOGY=========
\section{Methodology}
\label{methodology}

\subsection{A grouped panel GP regression model}
\label{methodology_subsect_1}

We first introduce a grouped panel GP regression model that parallels the GEV version proposed by \cite{Dupuis_2023}. Consider an $N \times T$ panel of observation $\{Y_{i,t}\}_{1\le i\le N,\;1\le t\le T}$, where each row corresponds to a series of observations for the same subject or location, and each column records measurements taken at the same time point or over the same time interval. The record length~$T$~is assumed to be much larger than the number of subjects/locations $N$, which is fixed. For each pair~$(i,t)$, we assume that a covariate vector $\mathbf{X}_{i,t}$ containing useful auxiliary information is available, which is treated as observed. For each~$Y_{i,t}$, we assume that for a high enough threshold $u_{i,t}$ the marginal distribution of the conditional excess $Z_{i,t}:= Y_{i,t} - u_{i,t} \, | \, Y_{i,t} > u_{i,t}$ can be well approximated by a GP distribution. We then focus on the panel of conditional excesses $\mathbf{Z}=\{Z_{i,t}:Y_{i,t}>u_{i,t}\}_{1\le i\le N,\;1\le t\le T}$, where observations below the thresholds are ignored but the relative positions of excesses in the original dataset are retained. For each $i \in \{1,\dots,N\}$, we denote the collection of time indices where exceedances are observed as~$\mathcal{T}_i$. That is, $\mathcal{T}_i:=\{t:Y_{i,t} > u_{i,t}, t=1,\ldots,T\}$. Now consider a latent partition of subjects denoted by $\mathbf{A} = (A_1, \dots, A_N)$ where $A_i \in \{1, \dots, \mathcal{G}\}$ for some integer $\mathcal{G}$ between 1 and~$N$. For each $Z_{i,t}\in\mathbf{Z}$ with $A_i = \tau\in\{1, \dots, \mathcal{G}\}$, we assume that its GP distributional parameters, transformed by link functions, linearly depend on the covariate vector $\mathbf{X}_{i,t}$ through some common regression parameters $\boldsymbol{\theta}_\tau=(\boldsymbol{\gamma}_{\tau}, \boldsymbol{\delta}_{\tau})$, where $\boldsymbol{\gamma}_{\tau}$ and $\boldsymbol{\delta}_{\tau}$ are $\sigma$-related and $\xi$-related regression parameters, respectively. That is, 
\begin{equation}
    Z_{i,t}|\mathbf{X}_{i,t}     
    = 
    \mathbf{x}_{i,t} \sim GP(\sigma_{i,t}(\tau), 
    \xi_{i,t}(\tau)),
\end{equation}
where
\begin{equation}
    \begin{split}
        \sigma_{i,t}(\tau) 
        &= e_{\sigma}(\boldsymbol{\gamma}_{\tau}^\top \mathbf{x}_{i,t}), \\
        \xi_{i,t}(\tau) &= e_{\xi}(\boldsymbol{\delta}_{\tau}^\top \mathbf{x}_{i,t}),
    \end{split}
\label{eqt:gp_GP_params}
\end{equation}
with link functions $e_{\sigma}(x) = e^x$ and $e_{\xi}(x) = x$. The covariates for the shape and scale parameters can differ, but we will not distinguish between them in the remainder of the paper. However, for subsequent use, we denote the dimensions of $\boldsymbol{\gamma}$ and $\boldsymbol{\delta}$ by dim($\boldsymbol{\gamma}$) and dim($\boldsymbol{\delta}$), respectively. Note that when the primary goal of inference is estimating return levels, the covariates used to explain temporal variation among extremes should be either stationary or exhibit a clear periodic pattern, with the length of the pattern being much smaller than the record length $T$, such that the observed covariate values are representative of their underlying distribution (\cite{eastoe2009modelling}). In many cases, one may consider using only time-invariant covariates instead (i.e., assuming only cross-sectional heterogeneity of extremes). In addition, although the shape parameters are allowed to be modelled as covariate-dependent, in practice it is common to assume that they are constant within each group. For notational consistency, we continue to use $\boldsymbol{\gamma}$ and $\boldsymbol{\delta}$ even when no covariates are incorporated. Among  all partitions that satisfy the group-wise GP
regression assumption above, we denote the one with the smallest value of $\mathcal{G}$ as $\mathbf{A}^*$ and implicitly assume its uniqueness for the sake of group structure identifiability. The number of groups under $\mathbf{A}^*$ is then denoted by~$\mathcal{G}^*$. Similar to \cite{Dupuis_2023}, we make no specific assumptions on the joint behaviour of the excesses. Instead, the only working assumption we make is that conditional on covariates, there exists a positive integer $m$ such that for any $Z_{i_1,t_1}$ and $Z_{i_2,t_2}$ where $|t_1-t_2|>m$, the two excesses are independent. In the sequel, we will refer to $m$ as the length of the largest dependence window. In practice, it should be determined by either domain knowledge or exploratory data analysis.

The estimation of $\mathbf{A}^*$ and $\mathcal{G}^*$ can still be done using the two-step maximization algorithm and the comparison of BIC values as suggested by \cite{Dupuis_2023}. Next we discuss the estimation of parameters and standard errors under the grouped panel GP regression model, assuming all subjects belong to a single group. The group index $\tau$ can thus be omitted.

\subsection{Parameter and covariance matrix estimation for a single group}
\label{methodology_subsect_2}

To estimate the regression parameters under the grouped panel GP regression model framework, we consider a maximum composite likelihood (MCL) estimator as in \cite{Dupuis_2023}, which they  refer to as a quasi maximum likelihood (QML) estimator. Here each component of the composite likelihood corresponds to an individual excess in
the panel data, and we restrict our attention to the unweighted case. Letting $\ell_{GP}(\boldsymbol{\theta}; z_{i,t}, \mathbf{x}_{i,t})$ denote the marginal log-likelihood evaluated at $\boldsymbol{\theta}$ for the excess $Z_{i,t}$ conditional on $\mathbf{X}_{i,t}$, the MCL estimator for the true parameter values $\boldsymbol{\theta}^*$ is defined as
\begin{equation}
    \begin{split} 
        \hat{\boldsymbol{ \theta}}^{ \text{GPR}}_{\text{MCL}} 
        &= \operatorname*{argmax}_{\boldsymbol{\theta}\in\mathbf{\Theta}}\sum^N_{i=1} \sum_{t\in\mathcal{T}_i} \ell_{GP}(\boldsymbol{\theta}; z_{i,t}, \mathbf{x}_{i,t})\\
        &= \operatorname*{argmax}_{(\boldsymbol{\gamma},\boldsymbol{\delta})\in\mathbf{\Theta}}\sum^N_{i=1} \sum_{t\in\mathcal{T}_i}\left[-\boldsymbol{\gamma}^\top\mathbf{x}_{i,t}-\left(\frac{1}{\boldsymbol{\delta}^\top\mathbf{x}_{i,t}}+1\right)\left(1+\frac{\boldsymbol{\delta}^\top\mathbf{x}_{i,t}z_{i,t}}{e^{\boldsymbol{\gamma}^\top\mathbf{x}_{i,t}}}\right)\right],
    \end{split}
\label{eqt:group_panel_GP_loglike}
\end{equation}
where $\mathbf{\Theta}$ is the restricted global parameter space subject to the constraint $1 + \frac{\boldsymbol{\delta}^\top \mathbf{x}_{i,t} z_{i,t}}{e^{\boldsymbol{\gamma}^\top \mathbf{x}_{i,t}}} > 0$. While studying the asymptotic properties of the MCL estimator above via verifying classical regularity conditions is complicated by the parameter dependence of the support of the GP distribution, we note that for asymptotic normality, the model regularity is in fact only required within a shrinking neighborhood of the true parameter values, where the chosen estimator lies with probability tending to one.  A relevant result in the context of the GEV distribution is given by \cite{bucher2017maximum}, and in the supplementary material we provide a similar result for the GP distribution.

Following our assumption on the dependence structure, the sandwich covariance estimator for $\hat{\boldsymbol{\theta}}^{\text{GPR}}_{\text{MCL}}$ should account for the dependence information across the entire dependence window, accumulating over time. Letting $\mathbf{S}_{GP}(\hat{\boldsymbol{\theta}}^{\text{GPR}}_{\text{MCL}}; z_{i,t}, \mathbf{x}_{i,t})$ denote the score vector evaluated at $\hat{\boldsymbol{\theta}}^{\text{GPR}}_{\text{MCL}}$ for the excess $Z_{i,t}$ conditional on $\mathbf{X}_{i,t}$, the sandwich estimator is given by ~$(\hat{{\mathbf{H}}}_{NT}^{\text{GPR}})^{-1}\hat{{\mathbf{V}}}^{\text{GPR}}_{NT}(\hat{{\mathbf{H}}}_{NT}^{\text{GPR}})^{-1}$, where
\begin{equation}
    \begin{split}           \hat{{\mathbf{V}}}^{\text{GPR}}_{NT}
    &= \sum^T_{t=1}    \left(\sum_{i:t\in\mathcal{T}_i}\mathbf{S}_{GP}(\hat{\boldsymbol{\theta}}^{\text{GPR}}_{\text{MCL}}; z_{i,t}, \mathbf{x}_{i,t})\right)\left(\sum_{i:t\in\mathcal{T}_i}\mathbf{S}_{GP}(\hat{\boldsymbol{\theta}}^{\text{GPR}}_{\text{MCL}}; z_{i,t}, \mathbf{x}_{i,t})\right)^\top \\
    &\quad + 2\sum^m_{l=1}\sum^T_{t=l+1}\left(\sum_{i:t-l\in\mathcal{T}_i}\mathbf{S}_{GP}(\hat{\boldsymbol{\theta}}^{\text{GPR}}_{\text{MCL}}; z_{i,t-l}, \mathbf{x}_{i,t-l})\right) \left(\sum_{i:t\in\mathcal{T}_i}\mathbf{S}_{GP}(\hat{\boldsymbol{\theta}}^{\text{GPR}}_{\text{MCL}}; z_{i,t}, \mathbf{x}_{i,t})\right)^\top,
    \end{split}
\label{eqt:GPR_MCL_V_hat}
\end{equation}
and 
\begin{equation}                                \hat{{ \mathbf{H}}}_{NT}^{ \text{GPR}}
    = \sum^N_{i=1}\sum_{t\in\mathcal{T}_i}\mathbf{S}_{GP}(\hat{\boldsymbol{\theta}}^{\text{GPR}}_{\text{MCL}}; z_{i,t}, \mathbf{x}_{i,t})\mathbf{S}_{GP}(\hat{\boldsymbol{\theta}}^{\text{GPR}}_{\text{MCL}}; z_{i,t}, \mathbf{x}_{i,t})^\top.
\label{eqt:GPR_MCL_H_hat}
\end{equation}
We examine the finite-sample performance of this dependence-window-based sandwich estimator under an extension of the grouped panel GP regression model, as introduced below.

\subsection{Extension to a two-level group structure}
\label{methodology_subsect_3}

In applications of the grouped panel GP and GEV regression models, subjects are grouped based on the similarities in both $\sigma$-related and $\xi$-related parameters. While this joint grouping enhances estimation efficiency, sometimes it may be too restrictive and make it difficult to pick a satisfactory model in practice. For example, when incorporating additional covariates for inference on the heterogeneity across the scale parameters, one should carefully address the potential bias introduced into the estimation of the $\xi$-related parameters. Moreover, the physical interpretation of the identified group structure can be vague, as the joint similarity of $\sigma$-related and $
\xi$-related parameters often results from a combination of various underlying factors in practice. The group structure based on the similarity of shape parameters themselves—which is typically difficult to determine a priori and may be of scientific interest, as it can potentially reveal the underlying factors driving the similarity in the tail behavior—cannot be analyzed in isolation.

To handle the challenges above, we introduce a two-level grouped panel GP regression model. While retaining the same regression setup as the grouped panel GP regression model in Section \ref{methodology_subsect_1}, which will be referred to as the one-level grouped version in the remainder of the paper, we now allow the $\sigma$-related and $
\xi$-related regression parameters to have different group structures. That is, consider the latent partitions
\begin{equation}                                \mathbf{A}_{\boldsymbol{\gamma}}
    = 
    (A_{\boldsymbol{\gamma},1}, \dots, A_{\boldsymbol{\gamma},N}), 
    \quad 
    A_{\boldsymbol{\gamma},i} \in \{1, \dots, \mathcal{G}_{\boldsymbol{\gamma}}\},
\label{eqt:double_group_assign_1}
\end{equation}
and
\begin{equation}
    \mathbf{A}_{\boldsymbol{\delta}}
    = 
    (A_{\boldsymbol{\delta},1}, \dots, A_{\boldsymbol{\delta},N}), 
    \quad 
    A_{\boldsymbol{\delta},i} \in \{1, \dots, \mathcal{G}_{\boldsymbol{\delta}}\},
\label{eqt:double_group_assign_2}
\end{equation}
where for each $Z_{i,t}$ with $A_{\boldsymbol{\gamma},i} = \tau_{\boldsymbol{\gamma}}$ and $A_{\boldsymbol{\delta},i} = \tau_{\boldsymbol{\delta}}$, we assume
\begin{equation}
    \begin{split}
    \sigma_{i,t}(\tau_{\boldsymbol{\gamma}}) &= e_{\sigma}(\boldsymbol{\gamma}_{\tau_{\boldsymbol{\gamma}}}^\top \mathbf{x}_{i,t}), \\
    \xi_{i,t}(\tau_{\boldsymbol{\delta}}) &= e_{\xi}(\boldsymbol{\delta}_{\tau_{\boldsymbol{\delta}}}^\top \mathbf{x}_{i,t}).
    \end{split}
\label{eqt:double_group_assign_reg_model}
\end{equation}
For group structure identifiability, similar to before, we implicitly assume that among all partition pairs satisfying (\ref{eqt:double_group_assign_reg_model}), there exists a unique partition pair $(\mathbf{A}_{\boldsymbol{\gamma}}^*, \mathbf{A}_{\boldsymbol{\delta}}^*)$ under group dimension pair $(\mathcal{G}_{\boldsymbol{\gamma}}^*, \mathcal{G}_{\boldsymbol{\delta}}^*)$ where for any possible group dimension pair $(\mathcal{G}_{\boldsymbol{\gamma}}, \mathcal{G}_{\boldsymbol{\delta}})$ we have $\mathcal{G}_{\boldsymbol{\gamma}}^*\leq\mathcal{G}_{\boldsymbol{\gamma}}$ and $\mathcal{G}_{\boldsymbol{\delta}}^*\leq\mathcal{G}_{\boldsymbol{\delta}}$. $(\mathbf{A}_{\boldsymbol{\gamma}}^*, \mathbf{A}_{\boldsymbol{\delta}}^*, \mathcal{G}_{\boldsymbol{\gamma}}^*, \mathcal{G}_{\boldsymbol{\delta}}^*)$ is then the target of group structure inference. 

\begin{figure}[!t]
    \centering    \includegraphics[width=9cm, height=8.5cm]{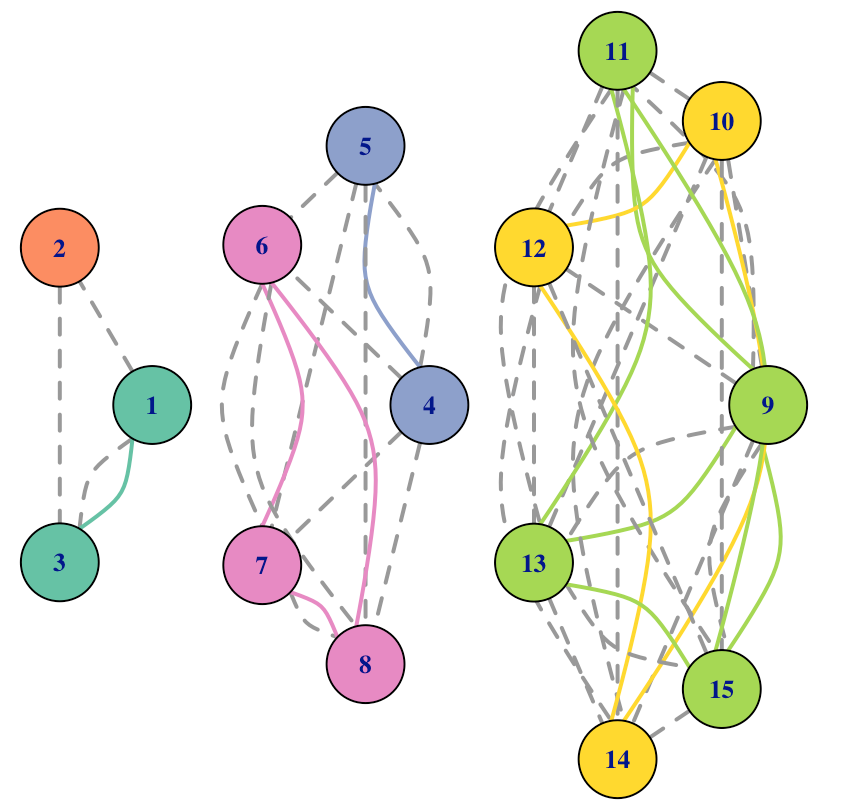}
    \captionsetup{
        format=hang,         
        width=1\textwidth, 
        font=small,          
        labelfont=bf,        
        justification=justified 
    }
    \caption{Three subject nets among 15 subjects are shown. Subjects are connected by dashed lines when they share the same $\sigma$-related parameters and by solid lines when they share the same $\xi$-related parameters. $\xi$-related parameter groups—reflecting the similarity in the tail behavior—are distinguished by color.}
    \label{fig:subj_nets}
\end{figure}

Compared to the one-level model, this two-level grouped extension allows exploration of a richer model space, where improved data fit does not necessarily come at the cost of increased model complexity. In fact, by decoupling the similarities in the $\sigma$-related and $\xi$-related regression parameters, we may identify models that simultaneously achieve greater parsimony and better fit. However, at the same time, inference under this model assumption requires more consideration. Let us first focus on the estimation of the parameters $(\boldsymbol{\gamma}_1,\dots, \boldsymbol{\gamma}_{\mathcal{G}_{\boldsymbol{\gamma}}}, \boldsymbol{\delta}_1,\dots, \boldsymbol{\delta}_{\mathcal{G}_{\boldsymbol{\delta}}^*})$ with the target group dimension pair $(\mathcal{G}_{\boldsymbol{\gamma}}^*, \mathcal{G}_{\boldsymbol{\delta}}^*)$ and the corresponding group structure $(\mathbf{A}_{\boldsymbol{\gamma}}^*, \mathbf{A}_{\boldsymbol{\delta}}^*)$ already known. As an adaptation of (\ref{eqt:group_panel_GP_loglike}) under the two-level grouped model setting, now the MCL estimator is defined with respect to all parameters associated with each parameter-linked subject net $\mathcal{C}\subseteq\{1,\dots,N\}$ that satisfies:
\begin{enumerate}
    \item 
    \textbf{Connectivity:} for any two subjects $i,j \in \mathcal{C}$, there exists a sequence of subjects $\{i=i_1,i_2,\dots,i_{k-1}, i_k=j: i_1,i_2,\dots,i_{k-1}, i_k \in \mathcal{C}\}$ such that for each pair $(i_m, i_{m+1})$, $m\leq k-1$, we have $\boldsymbol{\gamma}_{A^*_{\boldsymbol{\gamma}, i_m}}=\boldsymbol{\gamma}_{A^*_{\boldsymbol{\gamma}, i_{m+1}}}$ or $\boldsymbol{\delta}_{A^*_{\boldsymbol{\delta}, i_m}}=\boldsymbol{\delta}_{A^*_{\boldsymbol{\delta}, i_{m+1}}}$.
    \item 
    \textbf{Isolation:} for any two subjects $i \in \mathcal{C}$ and $j \not\in \mathcal{C}$, we have $\boldsymbol{\gamma}_{A^*_{\boldsymbol{\gamma}, i}}\neq\boldsymbol{\gamma}_{A^*_{\boldsymbol{\gamma}, j}}$ and $\boldsymbol{\delta}_{A^*_{\boldsymbol{\delta}, i}}\neq\boldsymbol{\delta}_{A^*_{\boldsymbol{\delta}, j}}$.
\end{enumerate}
Intuitively, a subject net $\mathcal{C}$ corresponds to a set of subjects connected via a parameter chain. Clearly any two distinct nets are disjoint, and the collection of all subject nets under a given partition $(\mathbf{A}_{\boldsymbol{\gamma}},\mathbf{A}_{\boldsymbol{\delta}})$ forms a partition of $\{1,\dots,N\}$. See Figure~\ref{fig:subj_nets} 
 for an illustrative example. For the justification of the use of MCL estimators in this case, see Theorem 2.1 of \cite{shao2023mle} for a related foundational result. 

We note that directly computing the MCL estimates over certain subject nets can be challenging due to high-dimensional optimization. For instance, consider the case where $\mathcal{C} = \{1, \dots, N\}$, $\mathcal{G}_{\boldsymbol{\gamma}}^* = \mathcal{G}_{\boldsymbol{\delta}}^* = 4$, with $\text{dim}(\boldsymbol{\gamma})=2$ and $\text{dim}(\boldsymbol{\delta})=1$. The parameter estimation leads to an optimization problem over a parameter vector in $\mathbb{R}^{12}$, which can pose notable numerical difficulties. We thus suggest an alternative strategy to compute the target estimates: for the vector of parameter estimates $(\hat{\boldsymbol{\gamma}}_1,\dots,\hat{\boldsymbol{\gamma}}_{\mathcal{G}_{\boldsymbol{\gamma}}^*},\hat{\boldsymbol{\delta}}_1,\dots,\hat{\boldsymbol{\delta}}_{\mathcal{G}_{\boldsymbol{\delta}}^*})$, we sequentially and cyclically update each of its components by maximizing a simpler partial composite log-likelihood associated with one component, while holding all the other components fixed at their current estimates, until convergence. Specifically, we randomly generate an initial value for each component and denote them by $(\hat{\boldsymbol{\gamma}}_1^{(0)},\dots,\hat{\boldsymbol{\gamma}}_{\mathcal{G}_{\boldsymbol{\gamma}}^*}^{(0)},\hat{\boldsymbol{\delta}}_1^{(0)},\dots,\hat{\boldsymbol{\delta}}_{\mathcal{G}_{\boldsymbol{\delta}}^*}^{(0)})$. Then, at each iteration $j\geq1$, for each $\tau_{\boldsymbol{\gamma}} \in \{1, \dots, \mathcal{G}^*_{\boldsymbol{\gamma}}\}$ and $\tau_{\boldsymbol{\delta}} \in \{1, \dots, \mathcal{G}^*_{\boldsymbol{\delta}}\}$, subject to the parameter constraints of the GP distribution, we update $\hat{\boldsymbol{\gamma}}_{\tau_{\boldsymbol{\gamma}}}^{(j)}$ and $\hat{\boldsymbol{\delta}}_{\tau_{\boldsymbol{\delta}}}^{(j)}$ by 
\begin{equation}
    \begin{split}       \hat{\boldsymbol{\gamma}}_{\tau_{\boldsymbol{\gamma}}}^{(j)} 
    &=
    \operatorname*{argmax}_{\boldsymbol{\gamma}}
    \sum_{i: A^*_{\boldsymbol{\gamma},i}=\tau_{\boldsymbol{\gamma}}} \sum_{t\in\mathcal{T}_i} \ell_{GP}(\boldsymbol{\gamma}, \hat{\boldsymbol{\delta}}^{(j-1)}_{A^*_{\boldsymbol{\delta},i}}; z_{i,t}, \mathbf{x}_{i,t})\\
    &= \operatorname*{argmax}_{\boldsymbol{\gamma}}\sum_{i: A^*_{\boldsymbol{\gamma},i}=\tau_{\boldsymbol{\gamma}}} \sum_{t\in\mathcal{T}_i}\Big[-\boldsymbol{\gamma}^\top\mathbf{x}_{i,t}-\Big(\frac{1}{(\hat{\boldsymbol{\delta}}^{(j-1)}_{A^*_{\boldsymbol{\delta},i}})^\top\mathbf{x}_{i,t}}+1\Big)\Big(1+\frac{(\hat{\boldsymbol{\delta}}^{(j-1)}_{A^*_{\boldsymbol{\delta},i}})^\top\mathbf{x}_{i,t}z_{i,t}}{e^{\boldsymbol{\gamma}^\top\mathbf{x}_{i,t}}}\Big)\Big],
    \end{split}
\label{eqt:multistep_gamma_update}
\end{equation}
and
\begin{equation}
    \begin{split}       \hat{\boldsymbol{\delta}}_{\tau_{\boldsymbol{\delta}}}^{(j)} 
    &=
    \operatorname*{argmax}_{\boldsymbol{\delta}}
    \sum_{i: A^*_{\boldsymbol{\delta},i}=\tau_{\boldsymbol{\delta}}} \sum_{t\in\mathcal{T}_i} \ell_{GP}(\hat{\boldsymbol{\gamma}}^{(j)}_{A^*_{\boldsymbol{\gamma},i}}, \boldsymbol{\delta}; z_{i,t}, \mathbf{x}_{i,t})\\
    &= \operatorname*{argmax}_{\boldsymbol{\delta}}\sum_{i: A^*_{\boldsymbol{\delta},i}=\tau_{\boldsymbol{\delta}}} \sum_{t\in\mathcal{T}_i}\Big[-(\hat{\boldsymbol{\gamma}}^{(j)}_{A^*_{\boldsymbol{\gamma},i}})^\top\mathbf{x}_{i,t}-\Big(\frac{1}{\boldsymbol{\delta}^\top\mathbf{x}_{i,t}}+1\Big)\Big(1+\frac{\boldsymbol{\delta}^\top\mathbf{x}_{i,t}z_{i,t}}{e^{(\hat{\boldsymbol{\gamma}}^{(j)}_{A^*_{\boldsymbol{\gamma},i}})^\top\mathbf{x}_{i,t}}}\Big)\Big],
    \end{split}
\label{eqt:multistep_delta_update}
\end{equation}
until all estimates converge numerically. Following \cite{tseng2001convergence}, such an optimization strategy can be viewed as an application of the block coordinate ascent (BCA) method, where in this case blocks are defined by $\{i: A^*_{\boldsymbol{\gamma},i}=\tau_{\boldsymbol{\gamma}}\}$ for $\tau_{\boldsymbol{\gamma}}=1,\dots, \mathcal{G}^*_{\boldsymbol{\gamma}}$ and $\{i: A^*_{\boldsymbol{\delta},i}=\tau_{\boldsymbol{\delta}}\}$ for $\tau_{\boldsymbol{\delta}}=1,\dots, \mathcal{G}^*_{\boldsymbol{\delta}}$. Note that the updates of parameter estimates across subject nets are independent, aligning with how the MCL estimators are defined. Moreover, conditional on the current values of $(\hat{\boldsymbol{\delta}}_1,\dots,\hat{\boldsymbol{\delta}}_{\mathcal{G}_{\boldsymbol{\delta}}^*})$, the updates to each component of $(\hat{\boldsymbol{\gamma}}_1,\dots,\hat{\boldsymbol{\gamma}}_{\mathcal{G}_{\boldsymbol{\gamma}}^*})$ are also independent, and they are expected to be much easier to perform than the joint optimization due to the reduced dimensionality in these smaller blocks relative to the full subject nets. The same also applies to estimating $(\hat{\boldsymbol{\gamma}}_1,\dots,\hat{\boldsymbol{\gamma}}_{\mathcal{G}_{\boldsymbol{\gamma}}^*})$ given~$(\hat{\boldsymbol{\delta}}_1,\dots,\hat{\boldsymbol{\delta}}_{\mathcal{G}_{\boldsymbol{\delta}}^*})$.

Despite the computational advantage of this strategy, when estimating the standard errors associated with each parameter estimate in
$(\hat{\boldsymbol{\gamma}}_1,\dots,\hat{\boldsymbol{\gamma}}_{\mathcal{G}_{\boldsymbol{\gamma}}^*},\hat{\boldsymbol{\delta}}_1,\dots,\hat{\boldsymbol{\delta}}_{\mathcal{G}_{\boldsymbol{\delta}}^*})$, one should construct sandwich covariance estimators at the subject-net level, incorporating all relevant parameter estimates within each net, rather than computing them separately within individual blocks. For the latter, taking $\hat{\boldsymbol{\gamma}}_{\tau_{\boldsymbol{\gamma}}}$ for illustration, one may consider first constructing a sandwich estimator over $(\hat{\boldsymbol{\gamma}}_{\tau_{\boldsymbol{\gamma}}},\{\hat{\boldsymbol{\delta}}_{A^*_{\boldsymbol{\delta},i}}:A^*_{\boldsymbol{\gamma},i}=\tau_{\boldsymbol{\gamma}}\})$ within the block defined by $\{i: A^*_{\boldsymbol{\gamma},i}=\tau_{\boldsymbol{\gamma}}\}$, and then extracting the dim($\boldsymbol{\gamma}$) $\times$ dim($\boldsymbol{\gamma}$) upper-left sub-matrix of the sandwich estimator as the estimated covariance matrix of $\hat{\boldsymbol{\gamma}}_{\tau_{\boldsymbol{\gamma}}}$. We note that this block-wise method yields only an approximate result in general because in this case the estimation efficiency of $\{\hat{\boldsymbol{\delta}}_{A^*_{\boldsymbol{\delta},i}}:A^*_{\boldsymbol{\gamma},i}=\tau_{\boldsymbol{\gamma}}\}$, whose impact on the standard errors of $\hat{\boldsymbol{\gamma}}_{\tau_{\boldsymbol{\gamma}}}$ is preserved through the non-zero off-diagonal elements of the covariance matrix of the parameter estimates associated with each subject net, is underestimated. Compared to the true variance of $\hat{\boldsymbol{\gamma}}_{\tau_{\boldsymbol{\gamma}}}$, the additional error may not be negligible. In contrast, despite the high dimensionality, constructing sandwich estimators at the subject-net level requires only the score vector and Hessian matrix evaluated at specific values, rather than full optimization. With covariates chosen appropriately, this typically does not result in numerical difficulties. Together with the suggested approach of computing the MCL estimates through the BCA strategy, the finite-sample performance of confidence intervals based on the normal approximation and the sandwich covariance estimator constructed at subject-net level is carefully evaluated through simulation studies, and satisfactory results are observed.

Another place where we may rely on the BCA strategy is the joint estimation of the group structure and parameters under a given group dimension pair $(\mathcal{G}_{\boldsymbol{\gamma}}, \mathcal{G}_{\boldsymbol{\delta}})$. Generalizing the two-step optimization algorithm proposed by \cite{Dupuis_2023} to the two-level grouped model setting with the incorporation of the BCA method, we introduce the following algorithm for the joint estimation:
\begin{enumerate}
    \item[] 
    \textbf{Step 1}: (\textit{Initialization}) Randomly generate initial group assignments of subjects:
    \begin{equation}
        \begin{split}            \hat{\mathbf{A}}^{(0)}_{\boldsymbol{\gamma}} 
        &= (\hat{A}^{(0)}_{\boldsymbol{\gamma},1}, \dots, \hat{A}^{(0)}_{\boldsymbol{\gamma},N}), \quad \hat{A}^{(0)}_{\boldsymbol{\gamma},i} \in \{1, \dots, \mathcal{G}_{\boldsymbol{\gamma}}\},\\    \hat{\mathbf{A}}^{(0)}_{\boldsymbol{\delta}} 
        &= (\hat{A}^{(0)}_{\boldsymbol{\delta},1}, \dots, \hat{A}^{(0)}_{\boldsymbol{\delta},N}), \quad \hat{A}^{(0)}_{\boldsymbol{\delta},i} \in \{1, \dots, \mathcal{G}_{\boldsymbol{\delta}}\},
        \end{split}
    \end{equation}
    for the $\sigma$-related and $\xi$-related regression parameters, respectively, and randomly generate
    \begin{equation}
        \begin{split}            \hat{\boldsymbol{\gamma}}^{(0)}
        &=(\hat{\boldsymbol{\gamma}}_1^{(0)},\dots,\hat{\boldsymbol{\gamma}}_{\mathcal{G}_{\boldsymbol{\gamma}}}^{(0)}),\\            \hat{\boldsymbol{\delta}}^{(0)}
        &=(\hat{\boldsymbol{\delta}}_1^{(0)},\dots,\hat{\boldsymbol{\delta}}_{\mathcal{G}_{\boldsymbol{\delta}}}^{(0)}),
        \end{split}
    \end{equation}
    as the initial parameter estimates under $\hat{\mathbf{A}}^{(0)}_{\boldsymbol{\gamma}}$ and $\hat{\mathbf{A}}^{(0)}_{\boldsymbol{\delta}}$.
    \item[]
    \textbf{Step 2}: (\textit{Multi-Step Maximization}) At each iteration $j\geq 1$, sequentially do the following steps until $\hat{\mathbf{A}}^{(j)}_{\boldsymbol{\gamma}}=\hat{\mathbf{A}}^{(j-1)}_{\boldsymbol{\gamma}}$, $\hat{\mathbf{A}}^{(j)}_{\boldsymbol{\delta}}=\hat{\mathbf{A}}^{(j-1)}_{\boldsymbol{\delta}}$, $\|\hat{\boldsymbol{\gamma}}^{(j)}-\hat{\boldsymbol{\gamma}}^{(j-1)}\|_\infty<\varepsilon$ and $\|\hat{\boldsymbol{\delta}}^{(j)}-\hat{\boldsymbol{\delta}}^{(j-1)}\|_\infty<\varepsilon$ for some small $\varepsilon$ selected to determine numerical convergence:
    \begin{itemize}
        \item[] (1) Update each component of $\hat{\boldsymbol{\gamma}}^{(j)}=(\hat{\boldsymbol{\gamma}}_1^{(j)},\dots,\hat{\boldsymbol{\gamma}}_{\mathcal{G}_{\boldsymbol{\gamma}}}^{(j)})$ as in (\ref{eqt:multistep_gamma_update}) but with $A_{\boldsymbol{\gamma},i}$ replaced by $ \hat{A}^{(j-1)}_{\boldsymbol{\gamma},i}$ and $G_{\boldsymbol{\gamma}}^*$ replaced by $G_{\boldsymbol{\gamma}}$.
        \item[] (2) Update each component of $\hat{\mathbf{A}}^{(j)}_{\boldsymbol{\gamma}}=(\hat{A}^{(j)}_{\boldsymbol{\gamma},1}, \dots, \hat{A}^{(j)}_{\boldsymbol{\gamma},N})$ by
        \begin{equation}        \hat{A}^{(j)}_{\boldsymbol{\gamma},i} = 
            \operatorname*{argmax}_{\tau_{\boldsymbol{\gamma}}\in\{1,\dots,\mathcal{G}_{\boldsymbol{\gamma}}\}}\sum_{t \in \mathcal{T}_i}\ell_{GP}( \hat{\boldsymbol{\gamma}}^{(j)}_{\tau_{\boldsymbol{\gamma}}},\hat{\boldsymbol{\delta}}^{(j-1)}_{\hat{A}^{(j-1)}_{\boldsymbol{\delta},i}}; z_{i,t}, \mathbf{x}_{i,t}), \quad i=1,\dots,N.
        \end{equation}
        \item[] (3) Update each component of $\hat{\boldsymbol{\delta}}^{(j)}=(\hat{\boldsymbol{\delta}}_1^{(j)},\dots,\hat{\boldsymbol{\delta}}_{\mathcal{G}_{\boldsymbol{\delta}}}^{(j)})$ as in (\ref{eqt:multistep_delta_update}) but with $A_{\boldsymbol{\delta},i}$ replaced by $ \hat{A}^{(j-1)}_{\boldsymbol{\delta},i}$ and $G_{\boldsymbol{\delta}}^*$ replaced by $G_{\boldsymbol{\delta}}$.
        \item[] (4) Update each component of $\hat{\mathbf{A}}^{(j)}_{\boldsymbol{\delta}}=(\hat{A}^{(j)}_{\boldsymbol{\delta},1}, \dots, \hat{A}^{(j)}_{\boldsymbol{\delta},N})$ by
        \begin{equation}     \hat{A}^{(j)}_{\boldsymbol{\delta},i} = 
            \operatorname*{argmax}_{\tau_{\boldsymbol{\delta}}\in\{1,\dots,\mathcal{G}_{\boldsymbol{\delta}}\}}\sum_{t \in \mathcal{T}_i}\ell_{GP}( \hat{\boldsymbol{\gamma}}^{(j)}_{\hat{A}^{(j)}_{\boldsymbol{\gamma},i}},\hat{\boldsymbol{\delta}}^{(j)}_{\tau_{\boldsymbol{\delta}}}; z_{i,t}, \mathbf{x}_{i,t}), \quad i=1,\dots,N.
        \end{equation}
    \end{itemize}
\end{enumerate}
We refer to this algorithm as the multi-step maximization algorithm, which can also be viewed as an EM-type algorithm. While it may be of interest to investigate how altering the order of the four maximization sub-steps in \textbf{Step 2} affects the algorithm’s efficiency in group structure identification, simulation studies suggest that the current ordering already yields satisfactory performance. For determining $(\hat{\mathcal{G}}_{\boldsymbol{\gamma}}^*, \hat{\mathcal{G}}_{\boldsymbol{\delta}}^*)$, it is still natural to consider the BIC-comparison method as suggested by \cite{Dupuis_2023}. That is, under the two-level grouped model setting, we select $(\hat{\mathcal{G}}_{\boldsymbol{\gamma}}^*, \hat{\mathcal{G}}_{\boldsymbol{\delta}}^*)$ by 
\begin{equation}                                 (\hat{\mathcal{G}}_{ 
    \boldsymbol{\gamma}}^*, 
    \hat{\mathcal{G}}_{ \boldsymbol{\delta}}^*) 
    = \operatorname*{argmin}_{(\mathcal{G}_{\boldsymbol{\gamma}}, \mathcal{G}_{\boldsymbol{\delta}})\in\{1,\dots,N\}\times\{1,\dots,N\}}\text{BIC}(\mathcal{G}_{\boldsymbol{\gamma}}, \mathcal{G}_{\boldsymbol{\delta}}),
\label{eqt:BIC-based-method-estimates}
\end{equation}
with
\begin{equation}
    \text{BIC}(\mathcal{G}_{\boldsymbol{\gamma}} \mathcal{G}_{\boldsymbol{\delta}}) = -2\sum^N_{i=1}\sum_{t\in\mathcal{T}_i}\ell_{GP}(\hat{\boldsymbol{\gamma}}_{\hat{A}_{\boldsymbol{\gamma},i}^*}, \hat{\boldsymbol{\delta}}_{\hat{A}_{\boldsymbol{\delta},i}^*}; z_{i,t},\mathbf{x}_{i,t})+\log\Big(\sum^N_{i=1}|\mathcal{T}_i|\Big)\Big(\text{dim}(\boldsymbol{\gamma})\cdot\mathcal{G}_{\boldsymbol{\gamma}}+\text{dim}(\boldsymbol{\delta})\cdot\mathcal{G}_{\boldsymbol{\delta}}\Big),
\label{eqt:BIC-based-method-BIC}
\end{equation}
where $\{\hat{\boldsymbol{\gamma}}_{\hat{A}_{\boldsymbol{\gamma},i}}:i=1,\dots,N\}$ and $\{\hat{\boldsymbol{\delta}}_{\hat{A}_{\boldsymbol{\delta},i}}:i=1,\dots,N\}$ are computed under $(\mathcal{G}_{\boldsymbol{\gamma}}, \mathcal{G}_{\boldsymbol{\delta}})$ using the multi-step maximization algorithm. Simulation studies show that overall the identification accuracy of this method improves as $T$ increases, agreeing with the pattern observed in the one-level grouped model setting. However, we note that some challenges associated with the group structure identification workflow above—as adapted from \cite{Dupuis_2023}—can become more pronounced under the two-level grouped model assumption. These issues motivate us to consider an alternative workflow for the group structure estimation, as introduced next.

\subsection{A two-stage approach relying on hierarchical clustering}
\label{methodology_subsect_4}

Like many other EM-type algorithms, the multi-step maximization algorithm can converge to suboptimal group structures. A default solution is to run the algorithm multiple times at each candidate group dimension pair and retain the one with the lowest BIC. However, under the two-level grouped model setting, the overall computational cost of such a method can be substantial. In addition to the need to explore various combinations of $\mathcal{G}_{\boldsymbol{\gamma}}$ and $\mathcal{G}_{\boldsymbol{\delta}}$, slow convergence of the multi-step maximization algorithm when the group dimensions take relatively large values is also observed. 

A key reason that leads to this challenge is that we aim to identify $\mathcal{G}_{\boldsymbol{\gamma}}^*$ and 
$\mathcal{G}_{\boldsymbol{\delta}}^*$ simultaneously. Note that the idea of using BIC to determine $(\mathcal{G}_{\boldsymbol{\gamma}}^*, \mathcal{G}_{\boldsymbol{\delta}}^*)$ builds on that, with suitable constraints satisfied, BIC is consistent in selecting one of the most parsimonious models among all two-level grouped panel GP models that non-identifiably generate the data. This also implies that, by fixing $\mathcal{G}_{\boldsymbol{\delta}}$ at a value larger than $\mathcal{G}_{\boldsymbol{\delta}}^*$, or $\mathcal{G}_{\boldsymbol{\gamma}}$ at a value larger than $\mathcal{G}_{\boldsymbol{\gamma}}^*$, BIC can consistently select a model with group dimension pair $(\mathcal{G}_{\boldsymbol{\gamma}}^*, \mathcal{G}_{\boldsymbol{\delta}})$ or $(\mathcal{G}_{\boldsymbol{\gamma}}, \mathcal{G}_{\boldsymbol{\delta}}^*)$, respectively, under the same constraints. These selected models share the same density function as the one under the group dimension pair $(\mathcal{G}_{\boldsymbol{\gamma}}^*, \mathcal{G}_{\boldsymbol{\delta}}^*)$, which is the target of inference. This motivates a two‐stage approach: first, we estimate $\hat{\mathcal{G}}_{\boldsymbol{\gamma}}^*$ while fixing $\mathcal{G}_{\boldsymbol{\delta}}$ at a value that is relatively large but smaller than $N$; then, with $\hat{\mathcal{G}}_{\boldsymbol{\gamma}}^*$ fixed, we estimate $\hat{\mathcal{G}}_{\boldsymbol{\delta}}^*$. This approach is expected to be less computationally intensive, as less exploration is required. Evidence from simulation studies suggests that, for both identifying $\mathcal{G}_{\boldsymbol{\gamma}}^*$ under a fixed $\mathcal{G}_{\boldsymbol{\delta}}$ and identifying $\mathcal{G}_{\boldsymbol{\delta}}^*$ under a fixed $\mathcal{G}_{\boldsymbol{\gamma}}$, the finite-sample performance of the BIC-comparison method improves as $T$ increases. However, the identification rates for $\mathcal{G}_{\boldsymbol{\delta}}^*$ tend to be notably lower than those for $\mathcal{G}_{\boldsymbol{\gamma}}^*$. This is not surprising, as identifying the similarity structure in the tail behavior is naturally more challenging. To achieve a better identification performance, assuming the shape parameters are all covariate-independent, we suggest a new workflow that incorporates the idea of hierarchical clustering to help determine $\mathcal{G}_{\boldsymbol{\delta}}^*$ (and $\mathbf{A}_{\boldsymbol{\delta}}^*$):
\begin{itemize}
    \item[] \textbf{Step 1:} 
    Choose a relatively large value for $\mathcal{G}_{\boldsymbol{\delta}}$ and fix it. Estimate the group structures and the corresponding parameters using the multi-step maximization algorithm across a sequence of values of $\mathcal{G}_{\boldsymbol{\gamma}}$, where $\mathcal{G}_{\boldsymbol{\gamma}}^*$ is believed to be one of them. Determine $\hat{\mathcal{G}}_{\boldsymbol{\gamma}}^*$ using BIC, and retain the corresponding $\hat{\mathbf{A}}_{\boldsymbol{\gamma}}^*$ and $\hat{\boldsymbol{\gamma}}$.
    \item[] \textbf{Step 2:} Fix the values of  $\hat{\mathcal{G}}^*_{\boldsymbol{\gamma}}$, $\hat{\mathbf{A}}_{\boldsymbol{\gamma}}^*$ and $\hat{\boldsymbol{\gamma}}$ from \textbf{Step 1} and denote the initial group assignment of subjects with respect to shape parameters by $\hat{\mathbf{A}}^{(0)}_{\boldsymbol{\delta}}=(1,\dots,N)$. Compute $\hat{\boldsymbol{\delta}}^{(0)}$ based on (\ref{eqt:multistep_delta_update}) and the subject-wise composite log-likelihoods $\boldsymbol{\ell}_{GP}^{(0)} = \Big(\ell_{GP}^{(0)}(\hat{\boldsymbol{\gamma}}_{\hat{\mathbf{A}}_{\boldsymbol{\gamma},1}^*}, \hat{\boldsymbol{\delta}}^{(0)}_1; \mathbf{z}_{1}, \mathbf{x}_{1}),\ldots,\ell_{GP}^{(0)}(\hat{\boldsymbol{\gamma}}_{\hat{\mathbf{A}}_{\boldsymbol{\gamma},N}^*}, \hat{\boldsymbol{\delta}}^{(0)}_N; \mathbf{z}_{N},\mathbf{z}_{N})\Big)$, where each subject is treated as a sub-cluster here. Do the following:
    \begin{itemize}
        \item[] (1) Set $convergence = FALSE$ and the iteration index $j=1$.
        \item[] (2) While $convergence = FALSE$ do:
        \begin{itemize}
            \item[] (2.1) For every pair of distinct sub-clusters $(\mathbf{c}_1, \mathbf{c}_2)$ in $\hat{\mathbf{A}}_{\boldsymbol{\delta}}^{(j-1)}$, compute the marginal BIC improvement by $\Delta_{\text{BIC}^{\text{marginal}}_{(\mathbf{c}_1, \mathbf{c}_2)}} = \text{BIC}^{\text{merged}}_{(\mathbf{c}_1, \mathbf{c}_2)}-\text{BIC}^{\text{composite}}_{(\mathbf{c}_1, \mathbf{c}_2)}$ with
            \begin{equation}
                \begin{split}          \text{BIC}^{\text{merged}}_{(\mathbf{c}_1, \mathbf{c}_2)}
                =                &-2\sum_{i\in\mathbf{c}_1\cup\mathbf{c}_2}    \sum_{t\in\mathcal{T}_i}\ell_{GP}(\hat{\boldsymbol{\gamma}}_{\hat{\mathbf{A}}_{\boldsymbol{\gamma},i}^*}, \hat{\boldsymbol{\delta}}_{\mathbf{c}_1\cup\mathbf{c}_2};z_{i,t},\mathbf{x}_{i,t})
                \\
                \quad&+\Big(\text{dim}(\boldsymbol{\gamma})\cdot|\{\hat{A}^*_{\boldsymbol{\gamma},i}:i\in\mathbf{c}_1\cup\mathbf{c}_2\}| +\text{dim}(\boldsymbol{\delta})\Big)\log\left(\sum_{i\in\mathbf{c}_1\cup\mathbf{c}_2}|\mathcal{T}_i|\right),
                \end{split}         \label{eqt:BIC_merged}
            \end{equation}
            and
            \begin{equation}
                \begin{split}        \text{BIC}^{\text{composite}}_{(\mathbf{c}_1, \mathbf{c}_2)}
                = &-2\Big[\sum_{i\in\mathbf{c}_1} \sum_{t\in\mathcal{T}_i}\ell_{GP}(\hat{\boldsymbol{\gamma}}_{\hat{\mathbf{A}}_{\boldsymbol{\gamma},i}^*}, \hat{\boldsymbol{\delta}}_{\mathbf{c}_1};z_{i,t},\mathbf{x}_{i,t})+\sum_{i\in\mathbf{c}_2} \sum_{t\in\mathcal{T}_i}\ell_{GP}(\hat{\boldsymbol{\gamma}}_{\hat{\mathbf{A}}_{\boldsymbol{\gamma},i}^*}, \hat{\boldsymbol{\delta}}_{\mathbf{c}_2};z_{i,t},\mathbf{x}_{i,t})\Big]\\
                \quad&+\Big(\text{dim}(\boldsymbol{\gamma})\cdot|\{\hat{A}^*_{\boldsymbol{\gamma},i}:i\in\mathbf{c}_1\cup\mathbf{c}_2\}| +\text{dim}(\boldsymbol{\delta})\cdot|\{\hat{A}^{(j-1)}_{\boldsymbol{\delta},i}:i\in\mathbf{c}_1\cup\mathbf{c}_2\}|\Big)\\
                \quad &\times \log\left(\sum_{i\in\mathbf{c}_1\cup\mathbf{c}_2}|\mathcal{T}_i|\right),
            \end{split}         \label{eqt:BICs_in_algo}
        \end{equation}
        where $\hat{\boldsymbol{\delta}}_{\mathbf{c}_1\cup\mathbf{c}_2}$ is the estimate of $\boldsymbol{\delta}$ using all excesses from sub-clusters $\mathbf{c}_1$ and $\mathbf{c}_2$, keeping $\boldsymbol{\gamma}$ fixed at $\hat{\boldsymbol{\gamma}}$.
        \item[] (2.2) Denote the smallest marginal BIC improvement across all sub-cluster pair at the current iteration by $\Delta^{j}_{\text{BIC}^{\text{marginal}}_{\text{min}}}$. If $\Delta^{j}_{\text{BIC}^{\text{marginal}}_{\text{min}}}<0$, then merge the two sub-clusters that correspond to this best improvement into one sub-cluster and update $\hat{\mathbf{A}}_{\boldsymbol{\delta}}^{(j-1)}\rightarrow \hat{\mathbf{A}}_{\boldsymbol{\delta}}^{(j)}$, $\boldsymbol{\ell}_{GP}^{(j-1)}\rightarrow\boldsymbol{\ell}_{GP}^{(j)}$ and $\hat{\boldsymbol{\delta}}^{(j-1)}\rightarrow \hat{\boldsymbol{\delta}}^{(j)}$. Update the iteration index to $j+1$. If $\Delta^{j}_{\text{BIC}^{\text{marginal}}_{\text{min}}}\geq0$, set $\hat{\mathcal{G}}_{\boldsymbol{\delta}}^*=|\{\hat{A}_{\boldsymbol{\delta},1}^{(j-1)},\dots,\hat{A}_{\boldsymbol{\delta},N}^{(j-1)}\}|$, $\hat{\mathbf{A}}_{\boldsymbol{\delta}}^* = \hat{\mathbf{A}}_{\boldsymbol{\delta}}^{(j-1)}$, $\hat{\boldsymbol{\delta}} = \hat{\boldsymbol{\delta}}^{(j-1)}$ and $convergence=TRUE$. 
        \end{itemize}
    \end{itemize}
    \item[] \textbf{Step 3:} Under the current values of $\hat{\mathcal{G}}_{\boldsymbol{\gamma}}^*$ and $\hat{\mathcal{G}}_{\boldsymbol{\delta}}^*$, update $\hat{\boldsymbol{\gamma}}$, $\hat{\boldsymbol{\delta}}$, $\hat{\mathbf{A}}_{\boldsymbol{\gamma}}^*$ and $\hat{\mathbf{A}}_{\boldsymbol{\delta}}^*$ using the multi-step maximization algorithm with $\hat{\mathbf{A}}_{\boldsymbol{\gamma}}^*$ and $\hat{\mathbf{A}}_{\boldsymbol{\delta}}^*$ from \textbf{Step 1} and \textbf{Step 2} as initial group assignments. Retain $\hat{\mathcal{G}}_{\boldsymbol{\gamma}}^*$, $\hat{\mathcal{G}}_{\boldsymbol{\delta}}^*$, and the values of $\hat{\mathbf{A}}_{\boldsymbol{\gamma}}^*$, $\hat{\mathbf{A}}_{\boldsymbol{\delta}}^*$, $\hat{\boldsymbol{\gamma}}$ and $\hat{\boldsymbol{\delta}}$ from the algorithm outputs as the final group assignments and parameter estimates. 
\end{itemize}
We make some remarks on the workflow above. The effectiveness of hierarchical clustering is supported by the richer subject-level information of extreme events under the POT framework. With bottom-up grouping, a good balance in model fit across subjects is also anticipated. Despite the large number of merging attempts required under the hierarchical grouping framework, within each attempt, the computation of parameter estimates is highly efficient. In addition, $\hat{\mathcal{G}}_{\boldsymbol{\delta}}^*$ and $\hat{\mathbf{A}}_{\boldsymbol{\delta}}^*$ are estimated simultaneously in \textbf{Step 2}, with no need to repeat this procedure multiple times as in the multi-step maximization algorithm; for the final run of the multi-step maximization algorithm in \textbf{Step 3}, which guarantees a local optimal solution, a fast convergence is expected under the good initialization provided by the outputs from \textbf{Step 1} and \textbf{Step 2}. Compared to the benchmark BIC-comparison method, we expect a notably higher accuracy in identifying $\hat{\mathcal{G}}_{\boldsymbol{\delta}}^*$ under the hierarchical clustering approach. One may consider implementing the hierarchical clustering procedure on the $\sigma$-related parameters as well, but then one needs to handle the non-identifiability of the marginal GP regression model at the initial stage of clustering due to time-invariant covariates. Potential solutions include adding extra constraints on parameter space and relying on an optimization algorithm that returns a local solution if there is no unique maximizer of the objective function. We apply the latter approach in the multi-step maximization algorithm to handle a similar problem.

In the remainder of the paper, we refer to the new workflow described above as the two-stage approach with hierarchical clustering. When $\hat{\mathcal{G}}_{\boldsymbol{\gamma}}^*$ and $\hat{\mathcal{G}}_{\boldsymbol{\delta}}^*$ are identified separately but still by selecting the one with the lowest BIC among all models identified by the multi-step maximization algorithm, we refer to it as the two-stage approach with BIC comparison. The procedure of identifying $\hat{\mathcal{G}}_{\boldsymbol{\gamma}}^*$ is the same in these two approaches. The workflow adapted from \cite{Dupuis_2023} is simply referred to as the original workflow. Next, we provide details of the simulation studies that examine the effectiveness of our methodology for inference. 

%=========SIMULATIONS=========
\section{Simulation Studies}
\label{Simulations}

All simulation studies are conducted under the two-level grouped panel GP regression model assumption. We first examine the performance of 
$\hat{\boldsymbol{\theta}}^{\text{GPR}}_{\text{MCL}}$ computed using the BCA strategy and the empirical coverage rates of confidence intervals based on the normal approximation and the dependence-window-based sandwich covariance estimator. The latent group structure is assumed to be known. Then, we evaluate the performance of the multi-step maximization algorithm in identifying the latent group structures with the target group dimension pair $(\mathcal{G}_{\boldsymbol{\gamma}}^*, \mathcal{G}_{\boldsymbol{\delta}}^*)$ provided, as well as the performance of the original workflow and the two-stage approach with BIC comparison in identifying $\mathcal{G}_{\boldsymbol{\gamma}}^*$ and $ \mathcal{G}_{\boldsymbol{\delta}}^*$. The performance of the two-stage approach with hierarchical clustering in identifying $ \mathcal{G}_{\boldsymbol{\delta}}^*$ is then evaluated for comparison.

\subsection{Data generating process}
\label{simulation_subsect_1}

Data are simulated as follows. For the panel of covariates $\{\mathbf{X}_{i,t}\}_{1\le i\le N,\;1\le t\le T}$, besides an intercept term, we let ${X}_{i,t}^{(1)}$ be the only element in each $\mathbf{X}_{i,t}$. For each subject, the evolution of $X^{(1)}_{i,t}$ depends on a subject-specific stationary AR(1) process, which has a fixed standard deviation of 1 and a mean of 0.5, and a noise that is cross-sectionally dependent  within each $\sigma$-related parameter group. Specifically, at time $t$, for each subject $i \in \{1,\dots,N\}$ where $A^*_i=\tau_{\boldsymbol{\gamma}} \in \{1,\ldots,\mathcal{G}^*_{\boldsymbol{\gamma}}\}$, we generate $X^{(1)}_{i,t}$ according to:
\begin{equation}
    \begin{split}
        X^{(1)}_{i,t} &= W_{i,t}+\epsilon_{i,t},\\
        W_{i,t} &= \phi_i W_{i, t-1} + \tilde{\epsilon}_{i,t}, \quad \tilde{\epsilon}_{i,t} \stackrel{i.i.d.}{\sim} N(0,1),\\ \boldsymbol{\epsilon}_{\tau, t} &= \{\epsilon_{i,t}: A^*_i = {\tau_{\boldsymbol{\gamma}}}\} \sim N(0, \boldsymbol{\Sigma}_{\tau_{\boldsymbol{\gamma}}}),
    \end{split}
\end{equation}
where $\phi_i \sim Unif(-0.5, 0.5)$ characterizes each individual-specific AR(1) process, and $\boldsymbol{\Sigma}_{\tau_{\boldsymbol{\gamma}}}$ characterizes the covariance of noises for subjects within the same $\sigma$-related parameter group. For different $t \in \{1,\ldots,T\}$, $\boldsymbol{\epsilon}_{\tau, t}$'s are independent. Then we generate a panel of excesses over high thresholds, denoted by $\mathbf{Z}=\{Z_{i,t}\}_{1\le i\le N,\;1\le t\le T}$, from a distribution with GP margins $\{H(\cdot;\sigma_{i,t}, \xi_{i,t})\}_{1\le i\le N,\;1\le t\le T}$, where a Gaussian copula $\mathbb{C}^{\text{gauss}}$ is used to capture the dependence between the margins, both temporally and cross-sectionally. Specifically, letting $\mathbf{U}=\{U_{i,t}\}_{1\le i\le N,\;1\le t\le T}$ be a $N\times T$ panel of probability integral transforms (PITs), we simulate excesses through
\begin{equation}
    \begin{split}
        \mathbf{U} 
        &= \{U_{i,t}\}_{1\le i\le N,\;1\le t\le T} \sim \mathbb{C}^{\text{gauss}}(\boldsymbol{\Sigma}_{NT}),\\
        \mathbf{Z} 
        &= \{H^{-1}(U_{i,t};\sigma_{i,t},\xi_{i,t})\}_{1\le i\le N,\;1\le t\le T,}
    \end{split}
\end{equation}
where $\boldsymbol{\Sigma}_{NT}$ controls the correlation structure of $\mathbf{U}$. Since each $Z_{i,t}$ is directly generated from a GP distribution, we choose a threshold of zero for all $Z_{i,t}$ across time and subjects; for each margin, we only allow the scale parameter to be covariate-dependent.

\subsection{Parameter estimation and confidence interval coverage}
\label{simulation_subsect_2}

\begin{figure}[!t]
    \centering
    \includegraphics[width=14cm, height=7cm]{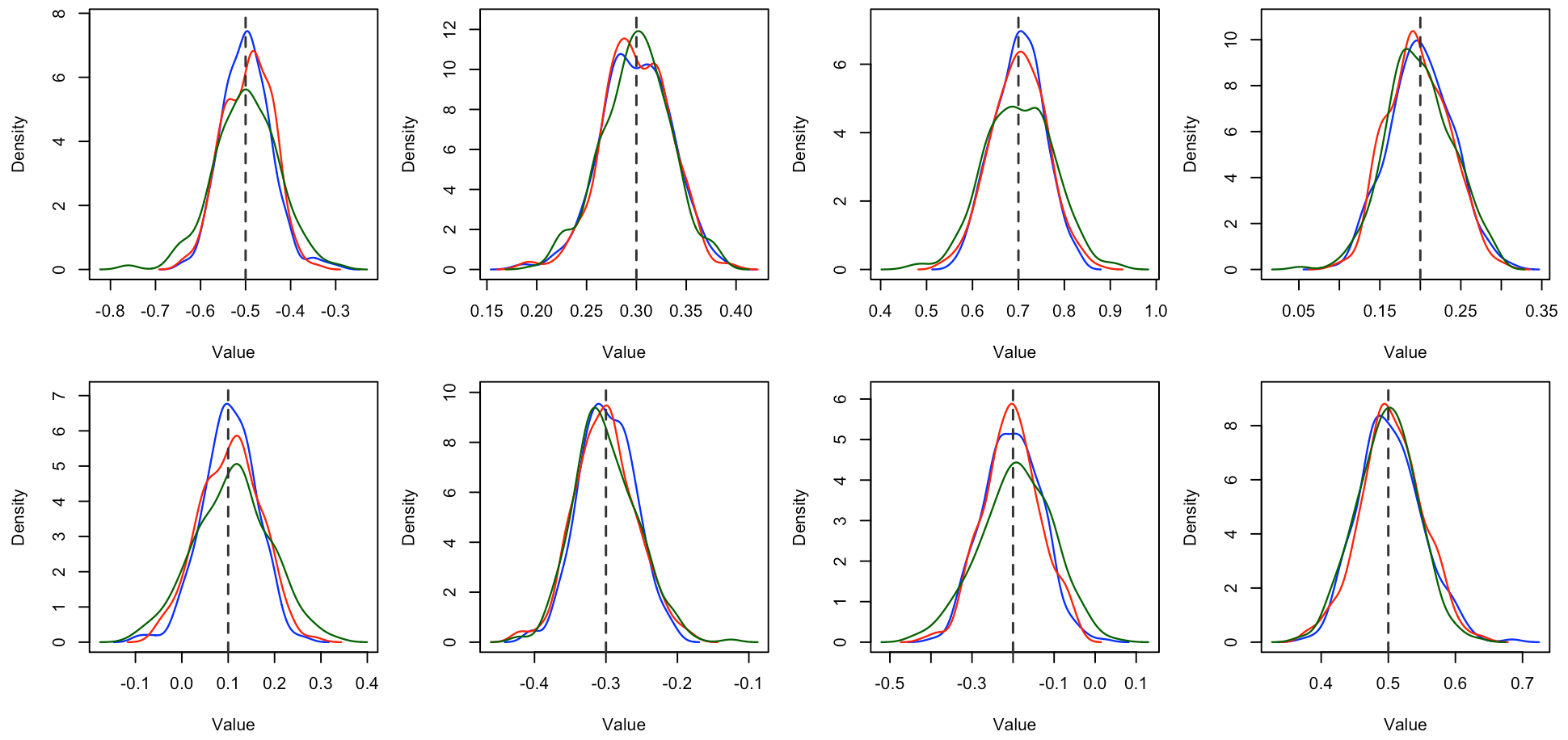}
    \includegraphics[width=14cm, height=5.6cm]{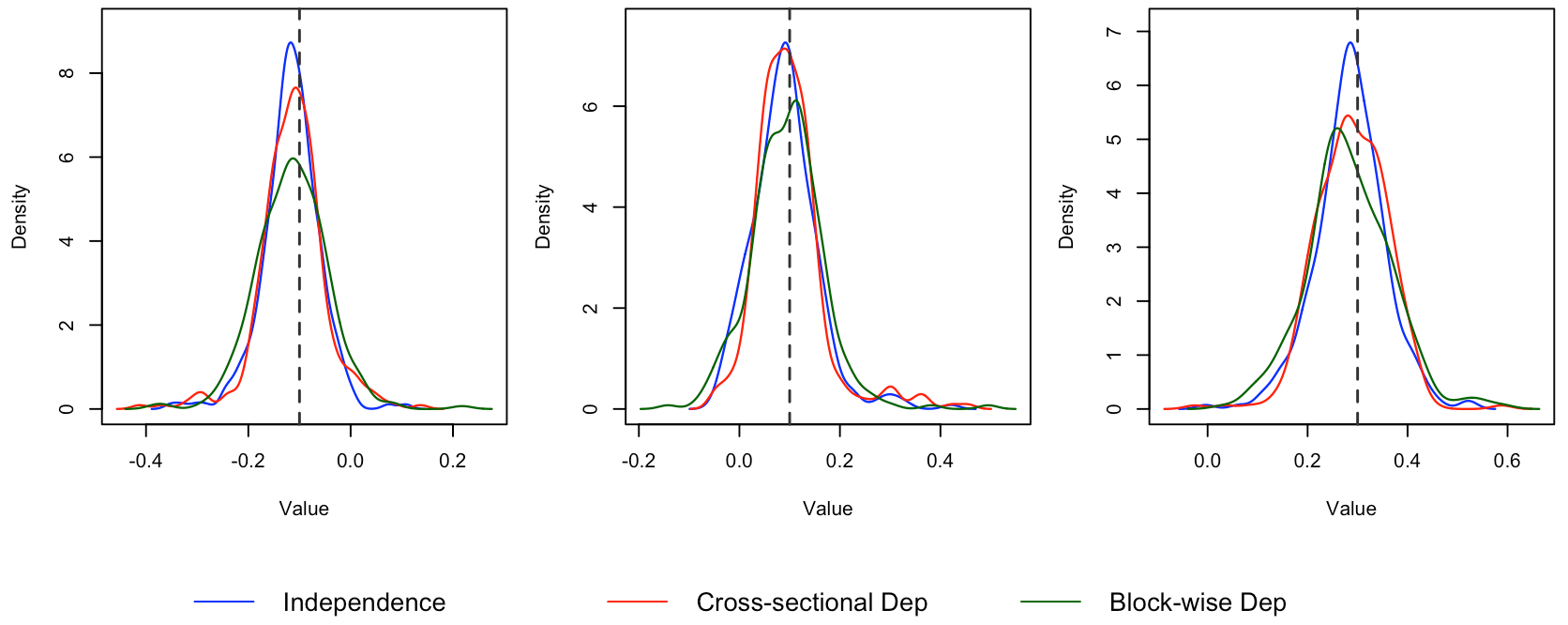}
    \caption{The sampling densities of the parameter estimates under three dependence structures for $T=2000$. The first two rows correspond to the eight parameters from the four $\sigma$-related parameter groups, and the third row corresponds to the three parameters from the three $\xi$-related parameter groups. The deep gray dashed lines indicate the true values of these parameters.}
    \label{fig:params_est_hists}
\end{figure}

We consider a scenario of 
12 subjects with
\begin{equation*}             \mathbf{A}^*_{\boldsymbol{\gamma}}=(1,1,1,2,2,2,3,3,3,4,4,4), \quad \mathbf{A}^*_{\boldsymbol{\delta}}=(1,1,1,1,2,2,2,2,3,3,3,3).
\end{equation*} 
That is, the 12 subjects form a single subject net as defined in Section \ref{methodology_subsect_3}. The two group structures are assumed to be known here. We consider three types of dependence structures (conditional on covariates):
\begin{enumerate} 
    \item \textbf{Independence}: All univariate excesses over thresholds are mutually independent. 
    \item \textbf{Cross-sectional dependence}: For each \( t \in \{1, \dots, T\} \), the excesses are cross-sectionally dependent, but they are independent across different time points.
    \item \textbf{Dependence within blocks of length \( m \)}: Suppose the panel of excesses \( \mathbf{Z} \) can be horizontally divided into \(T/m\) blocks, each of length \( m \). Within each block, the excesses are dependent, but they are independent across different blocks.
\end{enumerate}
\cite{Dupuis_2023} primarily focus on the second dependence structure. The third structure is a special case of the $m$-dependent process (with the first two structures also being $m$-dependent processes where $m = 0$). To replicate real-world scenarios where excesses tend to occur randomly, with only a subset of them exhibiting dependence, we generate data using moderate to large values of $T$ and then randomly sample small subsets from the panels for model fitting. Under this design, the third dependence structure is believed to serve as a reasonable approximation to real-world scenarios. We consider $T = 500, 1000,$ and $2000$ for all three dependence structures, with $10\%$ of the excesses randomly sampled for parameter estimation. Under the second and third dependence structures, the covariance matrix $ \boldsymbol{\Sigma}_{NT}$ is specified such that any pair of dependent excesses has a correlation of 0.9 if they belong to the same $\sigma$-related parameter group, and 0.1 if they do not. For the third dependence type, $m=4$ is chosen. For each sub-case, 300 replications are conducted, where at each replication we only generate the initial values for parameter estimates once and then apply the BCA method to obtain the MCL estimates. The parameter values used to generate the data are summarized in Table \ref{tab:true_param_values}
\begin{table}[!t]
    \centering
    \resizebox{\textwidth}{!}{ 
    \renewcommand{\arraystretch}{1.2}
    \begin{tabular}{
        >{\centering\arraybackslash}m{1.5cm}
        @{} >{\centering\arraybackslash}m{1.0cm} @{\hskip -0.8pt}
        >{\centering\arraybackslash}m{1.0cm} @{\hskip -0.8pt}
        >{\centering\arraybackslash}m{1.0cm} @{\hskip -0.8pt}
        >{\centering\arraybackslash}m{1.0cm} @{\hskip -0.8pt}
        >{\centering\arraybackslash}m{1.0cm} @{\hskip -0.8pt}
        >{\centering\arraybackslash}m{1.0cm} @{\hskip -0.8pt}
        >{\centering\arraybackslash}m{1.0cm} @{\hskip -0.8pt}
        >{\centering\arraybackslash}m{1.0cm} @{}
        >{\centering\arraybackslash}m{1.2cm}
        >{\centering\arraybackslash}m{1.2cm}
        >{\centering\arraybackslash}m{1.2cm}
    }
        \toprule
        \shortstack{} & 
        \multicolumn{8}{c}{$\sigma$-related parameters} & 
        \multicolumn{3}{c}{$\xi$-related parameters} \\
        \cmidrule(lr){2-9} \cmidrule(lr){10-12}
        \shortstack{Group} & 
        \multicolumn{2}{c}{$\tau_{\boldsymbol{\gamma}}=1$} & 
        \multicolumn{2}{c}{$\tau_{\boldsymbol{\gamma}}=2$} & 
        \multicolumn{2}{c}{$\tau_{\boldsymbol{\gamma}}=3$} & 
        \multicolumn{2}{c}{$\tau_{\boldsymbol{\gamma}}=4$} &
        $\tau_{\boldsymbol{\delta}}=1$ & 
        $\tau_{\boldsymbol{\delta}}=2$ & 
        $\tau_{\boldsymbol{\delta}}=3$ \\
        \cmidrule(lr){2-3} \cmidrule(lr){4-5} \cmidrule(lr){6-7} \cmidrule(lr){8-9}
        \cmidrule(lr){10-10} \cmidrule(lr){11-11} \cmidrule(lr){12-12}
        \shortstack{} & $\gamma_{1,0}$ & $\gamma_{1,1}$ & 
        $\gamma_{2,0}$ & $\gamma_{2,1}$ & 
        $\gamma_{3,0}$ & $\gamma_{3,1}$ & 
        $\gamma_{4,0}$ & $\gamma_{4,1}$ &
        $\delta_{1,0}$ & $\delta_{2,0}$ & $\delta_{3,0}$ \\
        \midrule
        Value & -0.5 & 0.3 & 0.7 & 0.2 & 0.1 & -0.3 & -0.2 & 0.5 & -0.1 & 0.1 & 0.3 \\
        \bottomrule
    \end{tabular}
    }
    \caption{Parameter values used to generate the data: each of the four $\sigma$-related groups includes two parameters, and each of the three $\xi$-related groups includes one parameter.}
    \label{tab:true_param_values}
\end{table}, and the simulation results for $T=2000$ are presented in Figure \ref{fig:params_est_hists}
.

In the case of $T=2000$, despite the reduced estimation efficiency under the cross-sectional and block-wise dependence structures, the density curves of the MCL estimates obtained using the BCA strategy remain approximately bell-shaped, with finite-sample bias and variance being only mildly to moderately worse than those under the independence case, agreeing with the claim of \cite{Chandler_Bate_2007}. In particularly, the estimation efficiency for the non-intercept regression parameters does not decline as the dependence strength increases. This is because when observations within the same cluster are highly dependent (conditional on covariates), but the covariates across these observations exhibit adequate variation, sometimes it is easier to estimate the impact of covariates on the distributional parameters of interest as the variation across observations driven by their mutual independence is reduced. For how the sampling densities of parameter estimates become increasingly concentrated around their true values and more bell-shaped as $T$ increases, the results under block-wise dependence are provided in the supplementary material. Together with those presented in Figure \ref{fig:params_est_hists}, the results support both the applicability of the MCL estimator under the two-level grouped model setting and the efficacy of the BCA strategy for computing the estimates. Next, we evaluate the empirical coverage rates of the approximate $95\%$ confidence intervals for the regression parameters based on the normal approximation and the proposed sandwich covariance estimator constructed over the whole subject net. For illustration, we once again focus exclusively on the coverage rates for $T=2000$, with the corresponding results displayed in Table \ref{tab:CI_coverage}
\begin{table}[!t]
\centering
\resizebox{\textwidth}{!}{ 
\begin{tabular}{lccccccccccc}
\toprule
 & \multicolumn{8}{c}{$\sigma$‐related parameters} 
 & \multicolumn{3}{c}{$\xi$‐related parameters} \\
\cmidrule(lr){2-9} \cmidrule(lr){10-12}
Dependence & 
  $\gamma_{1,0}$ & $\gamma_{1,1}$ & $\gamma_{2,0}$ & $\gamma_{2,1}$ & 
  $\gamma_{3,0}$ & $\gamma_{3,1}$ & $\gamma_{4,0}$ & $\gamma_{4,1}$ & 
  $\delta_{1,0}$ & $\delta_{2,0}$ & $\delta_{3,0}$ \\
\midrule
Independence & 0.97 & 0.95 & 0.95 & 0.94 & 0.97 & 0.97 & 0.95 & 0.95 & 0.95 & 0.94 & 0.94 \\
Cross‐sectional & 0.97 & 0.97 & 0.94 & 0.96 & 0.95 & 0.95 & 0.97 & 0.94 & 0.91 & 0.92 & 0.93 \\
Block‐wise & 0.94 & 0.93 & 0.94 & 0.92 & 0.93 & 0.94 & 0.94 & 0.97 & 0.90 & 0.94 & 0.93 \\
\bottomrule
\end{tabular}
}
\caption{Empirical coverage rates of approximate 95\% confidence intervals under the three dependence structures for $T=2000$, where 300 replications are conducted under each dependence structure.}
\label{tab:CI_coverage}
\end{table}. Across all three dependence structures, the coverage rates for the $\sigma$-related parameters remain close to the nominal coverage of $95\%$ with mild fluctuations. For $\delta_{1,0}$, $\delta_{2,0}$ and $\delta_{3,0}$ (i.e., the shape parameters themselves), it seems that coverage rates of the heavy-tailed group tend to be slightly higher than those of the short-tailed groups when the excesses are dependent. This is not surprising. For the short-tailed cases, as discussed by \cite{Davison1984}, constructing approximate confidence intervals based on asymptotic normality is usually worse than using other likelihood-based approaches, such as using profile likelihood, since the normal approximation here, as a first-order approximation, may deteriorate when the underlying distribution is short-tailed. Nevertheless, given the results above, we are comfortable to conclude that constructing confidence intervals using the normal approximation and the dependence-window-based sandwich covariance estimator provides satisfactory finite-sample performance under the two-level grouped panel GP regression model setting. The convergence rates based on standard errors estimated using the block-wise method are also examined, and the results are provided in the supplementary material.

\subsection{Group structure identification}
\label{simulation_subsect_3}

To examine the performance of the multi-step maximization algorithm in identifying the latent group structures with $(\mathcal{G}_{\boldsymbol{\gamma}}^*, \mathcal{G}_{\boldsymbol{\delta}}^*)$ given and the efficacy of the BIC-comparison approach in identifying~$(\mathcal{G}_{\boldsymbol{\gamma}}^*, \mathcal{G}_{\boldsymbol{\delta}}^*)$, we consider similar setups as above with a few modifications. Since cross-sectional dependence is a special case of block-wise dependence, we only consider the cases of independence and block-wise dependence with $m=4$ here. The within-same-$\sigma$-group correlation is set to be 0.5. For each sub-case, 200 replications are conducted, and we still sample only 10\% of excesses under each value of $T$.

Let us first focus on the identification of $\mathbf{A}_{\boldsymbol{\gamma}}^*$ and $\mathbf{A}_{\boldsymbol{\delta}}^*$ using the multi-step maximization algorithm with $(\mathcal{G}_{\boldsymbol{\gamma}}^*, \mathcal{G}_{\boldsymbol{\delta}}^*)=(4,3)$ provided. At each iteration, we only allow a single run of the algorithm. Although the two-stage approach with hierarchical clustering is primarily designed to improve the identification of $\mathcal{G}_{\boldsymbol{\delta}}^*$, as the identifications of $\mathbf{A}_{\boldsymbol{\gamma}}^*$ and $\mathbf{A}_{\boldsymbol{\delta}}^*$ are also affected, here we include the results under this approach for comparison. Specifically, we let the hierarchical clustering halt at $\mathcal{G}_{\boldsymbol{\delta}}=3$; in the \textbf{Step 1} of the two-stage approach, we still assume that $\mathcal{G}_{\boldsymbol{\delta}}^*$ is unknown. $\mathcal{G}_{\boldsymbol{\delta}}=\lceil\sqrt N \rceil$ is used as a rule of thumb, but alternative choices may also be considered. To measure the similarity between the true group structures and the estimated group structures, we use the Rand index (\cite{Rand_1971}) as the metric, which ranges from 0 to 1, with values closer to~1 indicating that the two group assignments are more similar. The results are displayed in Table \ref{tab:ri_comparison}
\begin{table}[!t]
\centering
\resizebox{\textwidth}{!}{ 
\begin{tabular}{llcccccc}
\toprule
 & \textbf{Dependence} 
   & \multicolumn{2}{c}{\(T=500\)} 
   & \multicolumn{2}{c}{\(T=1000\)} 
   & \multicolumn{2}{c}{\(T=2000\)} \\
\cmidrule(lr){3-4} \cmidrule(lr){5-6} \cmidrule(lr){7-8}
 & \textbf{Structure}
   & \textbf{Two-stage} & \textbf{Multi-step}
   & \textbf{Two-stage} & \textbf{Multi-step}
   & \textbf{Two-stage} & \textbf{Multi-step} \\
\midrule
\multirow{2}{*}{$\mathbf{A}_{\boldsymbol{\gamma}}^*$}
  & Independence & 0.88 & 0.89 & 0.93 & 0.93 & 0.96 & 0.96 \\
  & Block-Dep  & 0.89 & 0.89 & 0.94 & 0.93 & 0.95 & 0.94 \\
\addlinespace
\multirow{2}{*}{$\mathbf{A}_{\boldsymbol{\delta}}^*$}
  & Independence & 0.66 & 0.68 & 0.75 & 0.74 & 0.82 & 0.80 \\
  & Block-Dep  & 0.66 & 0.68 & 0.74 & 0.76 & 0.83 & 0.81 \\
\bottomrule
\end{tabular}
}
\caption{Average Rand index between the true $\mathbf{A}_{\boldsymbol{\gamma}}^*$ and $\mathbf{A}_{\boldsymbol{\delta}}^*$ and the group structures estimated using the two-stage approach with hierarchical clustering and solely the multi-step algorithm under two dependence structures and $T=500, 1000$, and $2000$. Across the 200 replications, a very small fraction was discarded due to algorithm failures stemming from the initial group assignments. All results are rounded to two decimal places.}
\label{tab:ri_comparison}
\end{table}. As expected, the performance of both methods improves as \(T\) increases. The dependence structure has only minor impact on the results. The identification performance for $\mathbf{A}_{\boldsymbol{\gamma}}^*$ consistently outpaces that for $\mathbf{A}_{\boldsymbol{\delta}}^*$, in line with our earlier discussion. However, given the difficulty of tail estimation and that all average Rand indices exceed 0.8 at \(T=2000\), it remains reasonable to conclude that both methods deliver satisfactory finite‐sample performances, with the two-stage approach with the hierarchical clustering exhibiting a slight edge. 

\begin{figure}[!t]
    \centering
    \begin{minipage}[t]{0.3\textwidth}
        \centering        {\includegraphics[width=0.98\linewidth]{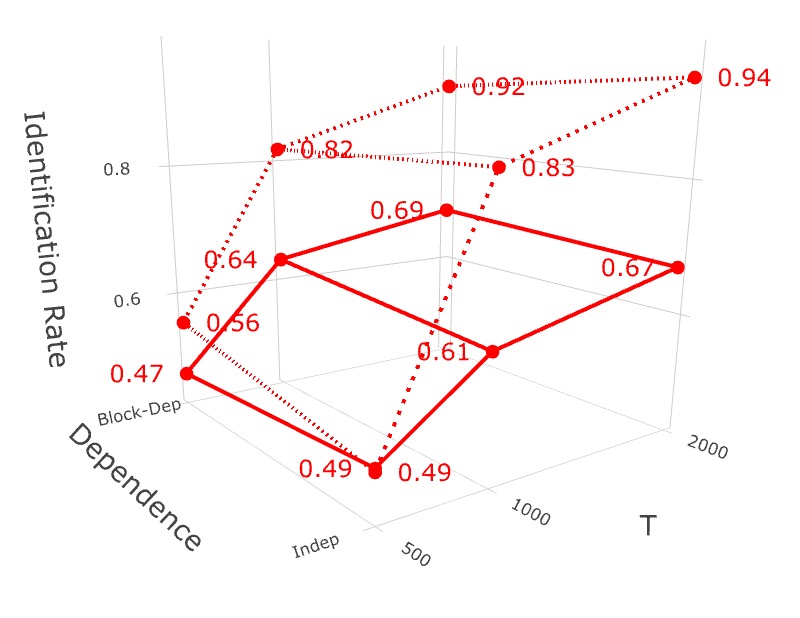}}
        \subcaption*{\parbox[t]{0.9\linewidth}{(a) Performance in identifying $\mathcal{G}^*_{\boldsymbol{\gamma}}$ given $\mathcal{G}_{\boldsymbol{\delta}}=2$}}
    \end{minipage}
    \begin{minipage}[t]{0.3\textwidth}
        \centering        {\includegraphics[width=0.98\linewidth]{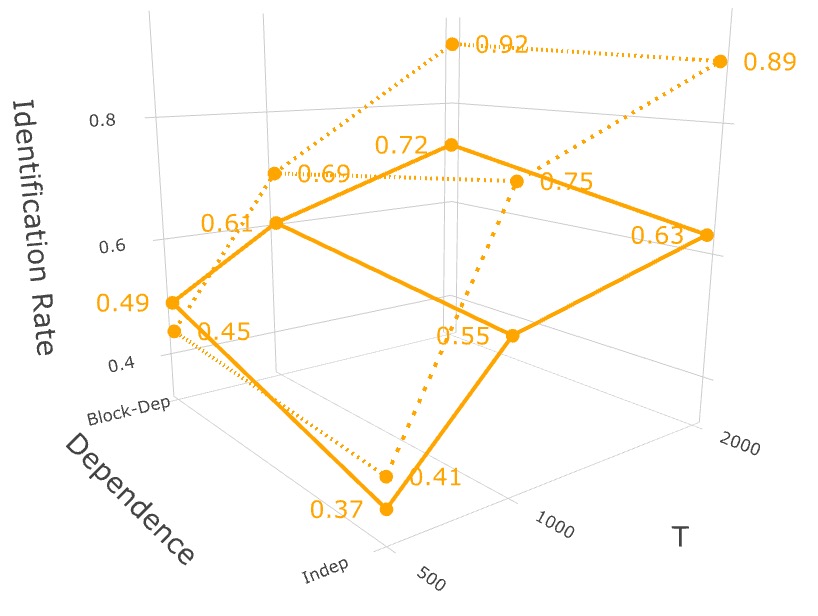}}
        \subcaption*{\parbox[t]{0.9\linewidth}{(b) Performance in identifying $\mathcal{G}^*_{\boldsymbol{\gamma}}$ given $\mathcal{G}_{\boldsymbol{\delta}}=3$}}
    \end{minipage}
    \begin{minipage}[t]{0.3\textwidth}
        \centering        {\includegraphics[width=0.98\linewidth]{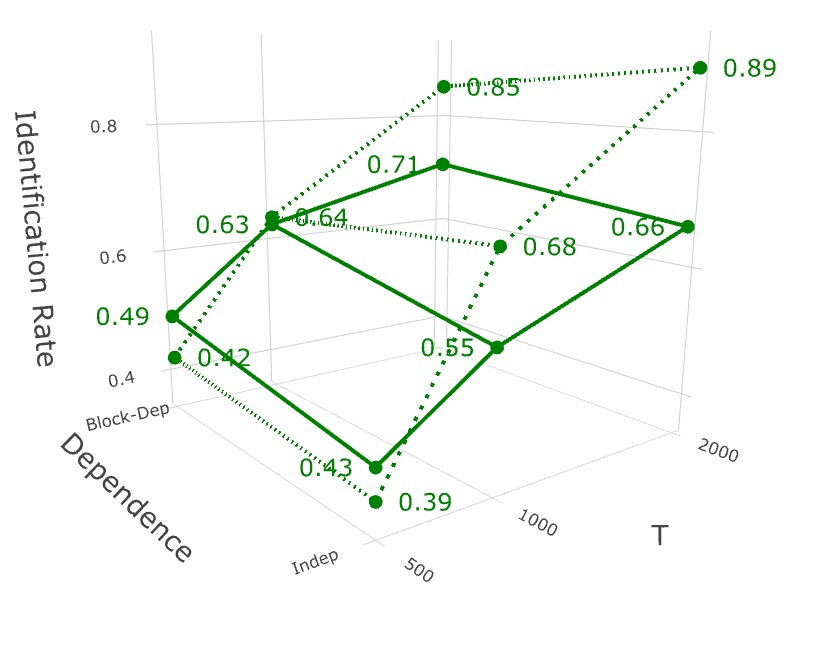}}
        \subcaption*{\parbox[t]{0.9\linewidth}{(c) Performance in identifying $\mathcal{G}^*_{\boldsymbol{\gamma}}$ given $\mathcal{G}_{\boldsymbol{\delta}}=4$}}
    \end{minipage}
    \vspace{0.5cm}
    \begin{minipage}[t]{0.3\textwidth}
        \centering        {\includegraphics[width=0.98\linewidth]{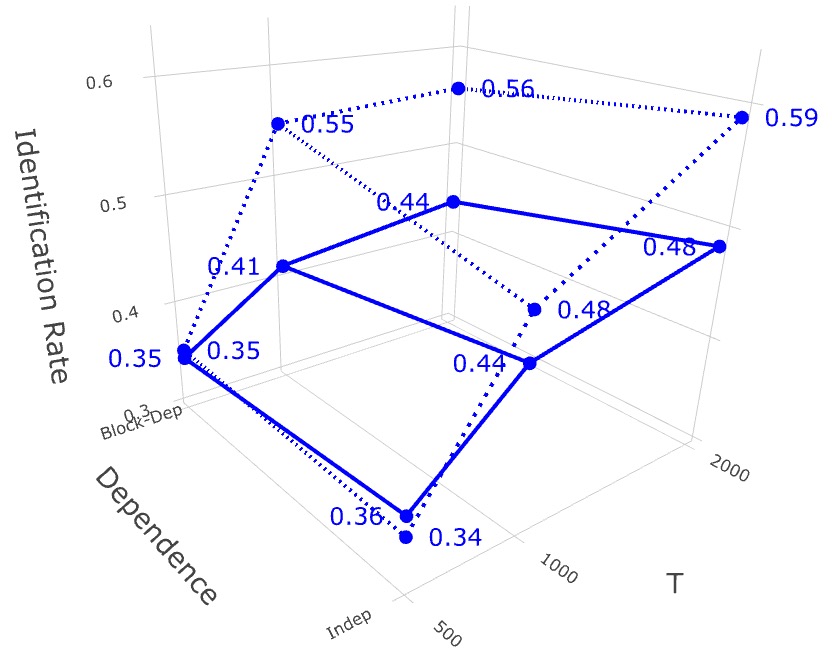}}
        \subcaption*{\parbox[t]{0.9\linewidth}{(d) Performance in identifying $\mathcal{G}^*_{\boldsymbol{\delta}}$ given $\mathcal{G}_{\boldsymbol{\gamma}}=3$}}
    \end{minipage}
    \begin{minipage}[t]{0.3\textwidth}
        \centering        {\includegraphics[width=0.98\linewidth]{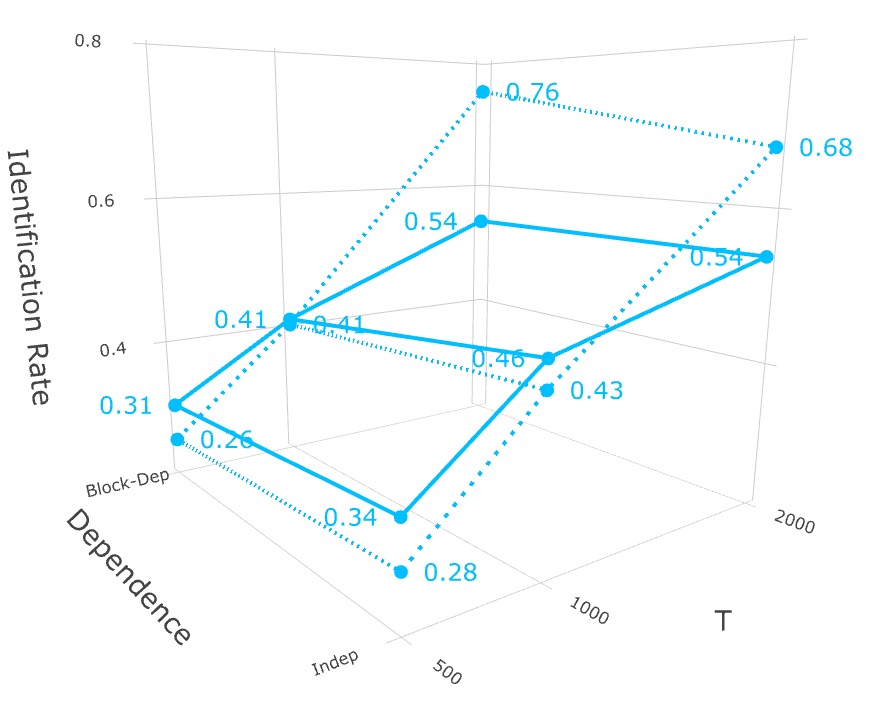}}
        \subcaption*{\parbox[t]{0.9\linewidth}{(e) Performance in identifying $\mathcal{G}^*_{\boldsymbol{\delta}}$ given $\mathcal{G}_{\boldsymbol{\gamma}}=4$}}
    \end{minipage}
    \begin{minipage}[t]{0.3\textwidth}
        \centering        {\includegraphics[width=0.98\linewidth]{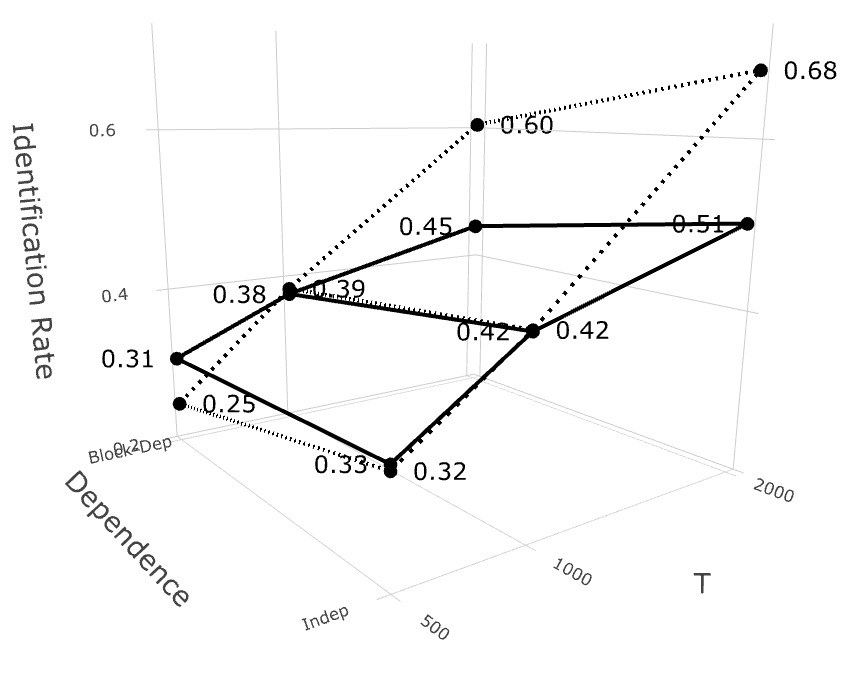}}
        \subcaption*{\parbox[t]{0.9\linewidth}{(f) Performance in identifying $\mathcal{G}^*_{\boldsymbol{\delta}}$ given $\mathcal{G}_{\boldsymbol{\gamma}}=5$}}
    \end{minipage}
    \begin{minipage}[t]{0.4\textwidth}
        \centering        {\includegraphics[width=0.98\linewidth]{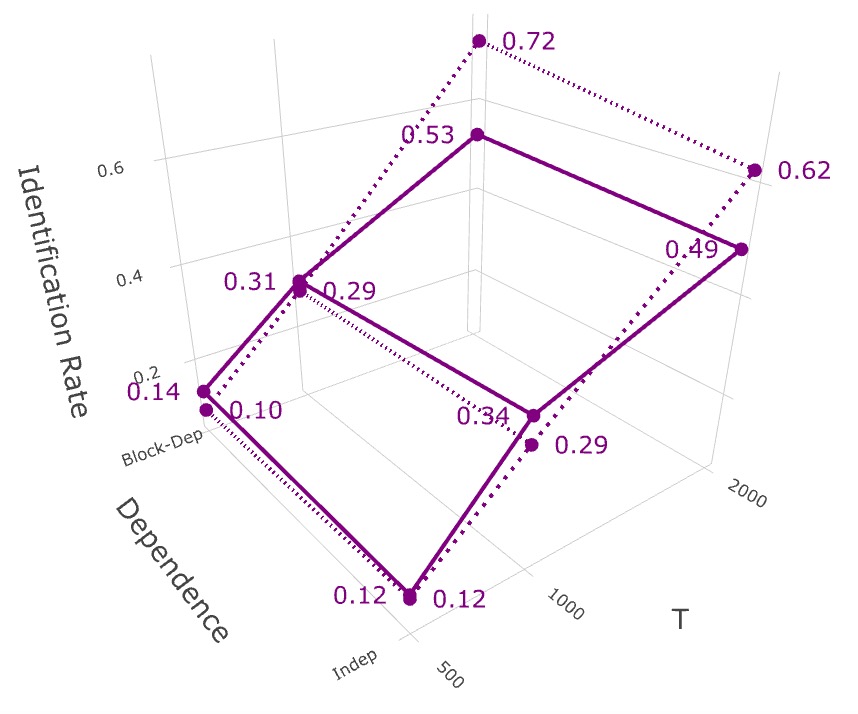}}
        \subcaption*{\parbox[t]{\linewidth}{(g) Performance in jointly identifying $\mathcal{G}^*_{\boldsymbol{\gamma}}$ and $\mathcal{G}^*_{\boldsymbol{\delta}}$}}
    \end{minipage}
    \caption{Joint and conditional group dimension(s) identification performance based on BIC comparison across different group dimension pairs, under two dependence structures and $T\in\{500, 1000, 2000\}$. Dashed lines correspond to the setting where the multi-step maximization algorithm is run three times per group dimension pair and the one with the lowest BIC is retained for further comparison. Solid lines correspond to the setting where the algorithm is run only once.}
    \label{fig:identification-rates}
\end{figure}

Next we examine the finite-sample performance of using BIC to jointly and separately identify $\mathcal{G}_{\boldsymbol{\gamma}}^*$ and $ \mathcal{G}_{\boldsymbol{\delta}}^*$. At each replication, we compare the BIC values of the group structures estimated by the multi-step maximization algorithm under the group dimension pairs $(3,2)$, $(3,3)$, $(3,4)$,  $(4,2)$, $(4,3)$, $(4,4)$, $(5,2)$, $(5,3)$ and $(5,4)$. If the BIC at $(\mathcal{G}^*_{\boldsymbol{\gamma}}, \mathcal{G}^*_{\boldsymbol{\delta}}) = (4,3)$, at $\mathcal{G}^*_{\boldsymbol{\gamma}} = 4$ under a given $\mathcal{G}_{\boldsymbol{\delta}}$, or at $\mathcal{G}^*_{\boldsymbol{\delta}} = 3$ under a given $\mathcal{G}_{\boldsymbol{\gamma}}$ is the smallest among all respective candidates, we say that the joint (or conditional) group dimension(s) are successfully identified. The results are given in Figure \ref{fig:identification-rates} 
. We summarize the results as follows. First, as we expect, the joint identification rates improve as $T$ increases, and similar patterns are observed for the conditional identification rates, even when conditional on group dimensions that are smaller than the true dimensions. However, no general conclusion can be drawn regarding the effect of the dependence structure. Second, allowing the multi-step maximization algorithm to run three times per group dimension pair leads to a notable improvement in identification rates when $T$ is large but yields little difference when $T$ is relatively small, likely due to the poor performance of BIC itself at smaller sample sizes. Although the former supports the choice of using BIC for model comparison under the two-level grouped panel GP regression model assumption, as previously discussed, we may still consider determining $\mathcal{G}^*_{\boldsymbol{\gamma}}$ and $\mathcal{G}^*_{\boldsymbol{\delta}}$ sequentially due to computational consideration. However, despite a good performance in identifying $\mathcal{G}^*_{\boldsymbol{\gamma}}$, we notice that even under three algorithm runs per group dimension pair, the conditional identification performance for $\mathcal{G}^*_{\boldsymbol{\delta}}$ is still not very satisfactory. Concerns are thus raised when applying the method to real-world applications involving more complex group structures, where a substantial number of algorithm runs may be required to accurately recover both $\mathcal{G}^*_{\boldsymbol{\delta}}$ and $\mathbf{A}^*_{\boldsymbol{\delta}}$. This directly motivates the new workflow proposed in Section~\ref{methodology_subsect_4}, and here we provide a comparison of the identification performance for $\mathcal{G}^*_{\boldsymbol{\delta}}$ (with $\mathcal{G}^*_{\boldsymbol{\gamma}} = 4$ provided) between the two-stage approach with BIC comparison (i.e., the results shown in panel~(e) of Figure~\ref{fig:identification-rates}) and the two-stage approach with hierarchical clustering. The results are summarized in Table \ref{tab:G_shape_identification}
\begin{table}[!t]
\centering
\resizebox{1.05\textwidth}{!}{
\begin{tabular}{llcccccc}
\toprule
 & \textbf{Dependence} 
   & \multicolumn{2}{c}{\(T=500\)} 
   & \multicolumn{2}{c}{\(T=1000\)} 
   & \multicolumn{2}{c}{\(T=2000\)} \\
\cmidrule(lr){3-4} \cmidrule(lr){5-6} \cmidrule(lr){7-8}
 & \textbf{Structure}
   & \textbf{Hierarchical} & \textbf{BIC Comparison}
   & \textbf{Hierarchical} & \textbf{BIC Comparison}
   & \textbf{Hierarchical} & \textbf{BIC Comparison} \\
\midrule
\multirow{2}{*}{$\mathcal{G}_{\boldsymbol{\delta}}^*$}
  & Independence & 0.48 & 0.28 & 0.63 & 0.43 & 0.77 & 0.68 \\
  & Block-Dep  & 0.49 & 0.26 & 0.56 & 0.41 & 0.78 & 0.76 \\
\bottomrule
\end{tabular}
}
\caption{The identification rates for $\mathcal{G}_{\boldsymbol{\delta}}^*$ conditional on $\mathcal{G}_{\boldsymbol{\gamma}}^*=4$ under the two-stage approach with hierarchical clustering and the two-stage approach with BIC comparison. At each group dimension pair the multi-step algorithm is run three times, under two dependence structures and $T=500, 1000$ and $2000$.}
\label{tab:G_shape_identification}
\end{table}. Across the two dependence structures and the three values of $T$, the performance of hierarchical clustering is consistently better than that of direct BIC comparison, particularly when $T$ is relatively small. Given this observation, along with the inconvenience of repeatedly running the multi-step maximization algorithm across a sequence of $\mathcal{G}_{\boldsymbol{\delta}}$ values—where slow convergence may occur—we recommend directly using the two-stage approach with hierarchical clustering when implementing the two-level grouped panel GP regression model in practice.

Next, we illustrate the application of our methodology using a real-world example.

%=========APPLICATION=========
\section{Application to river flow data of the upper Danube basin}
\label{Application}

Estimating extreme quantiles of river flow has long been one of the key application areas of extreme value theory. In many hydrological applications, river flow data from multiple stations are available; while the naive method of fitting a local GP model to the excesses over a high threshold for each station is a common choice, our methodology may provide a more efficient estimation of flood risk for each station by identifying the optimal group structure among a broader model space. For illustration, we apply our methodology to the daily river flow data of 31 hydrological stations in the upper Danube river basin, spanning 50 years from 1960 to 2009. This dataset has been frequently used in recent studies (see \cite{hentschel2024statistical}, \cite{Dupuis_2023}, \cite{engelke2020graphical} and \cite{asadi2015extremes}, among others), most of which concentrate on analyzing river network structures and/or identifying group patterns among different stations. Following \cite{asadi2015extremes} and \cite{engelke2020graphical}, we focus exclusively on the river flow records in June, July, and August, where the effects of seasonality can be ignored. In addition, as noted in \cite{asadi2015extremes}, national studies indicate that there are no notable trends in the discharge data across the 31 stations. This justifies estimating the return levels at each of these stations. 

The panel of river flow data has dimensions of $T = 4600$ ($92$ days/year $\times$ $50$ years) and $N = 31$ (stations). For each station, it is reasonable to assume that the summer river flow data over the 50 years constitutes a representative sample of their true distribution, at least within a specific period of interest. Available characteristics of each station include its longitude, latitude, average altitude, average volume, watershed area and mean slope of the corresponding subcatchment, which are all time-invariant if used as covariates to explain the cross-sectional heterogeneity of extremes. Daily precipitation records at these stations (or nearby stations that are not intended for river flow recording) are either unavailable or contain a significant number of missing values, and therefore they are excluded from the analysis. As a common choice in practice, we use the 0.95 quantile of the local river flow data as the high threshold to define exceedances at each station. We set the dependence window width to 9 days for our later construction of the sandwich covariance estimators, which corresponds to the maximum duration of an extreme flood event identified by \cite{asadi2015extremes}. 

\begin{figure}[!t]
    \centering
    \begin{minipage}[t]{0.49\textwidth}
        \centering        {\includegraphics[width=\linewidth]{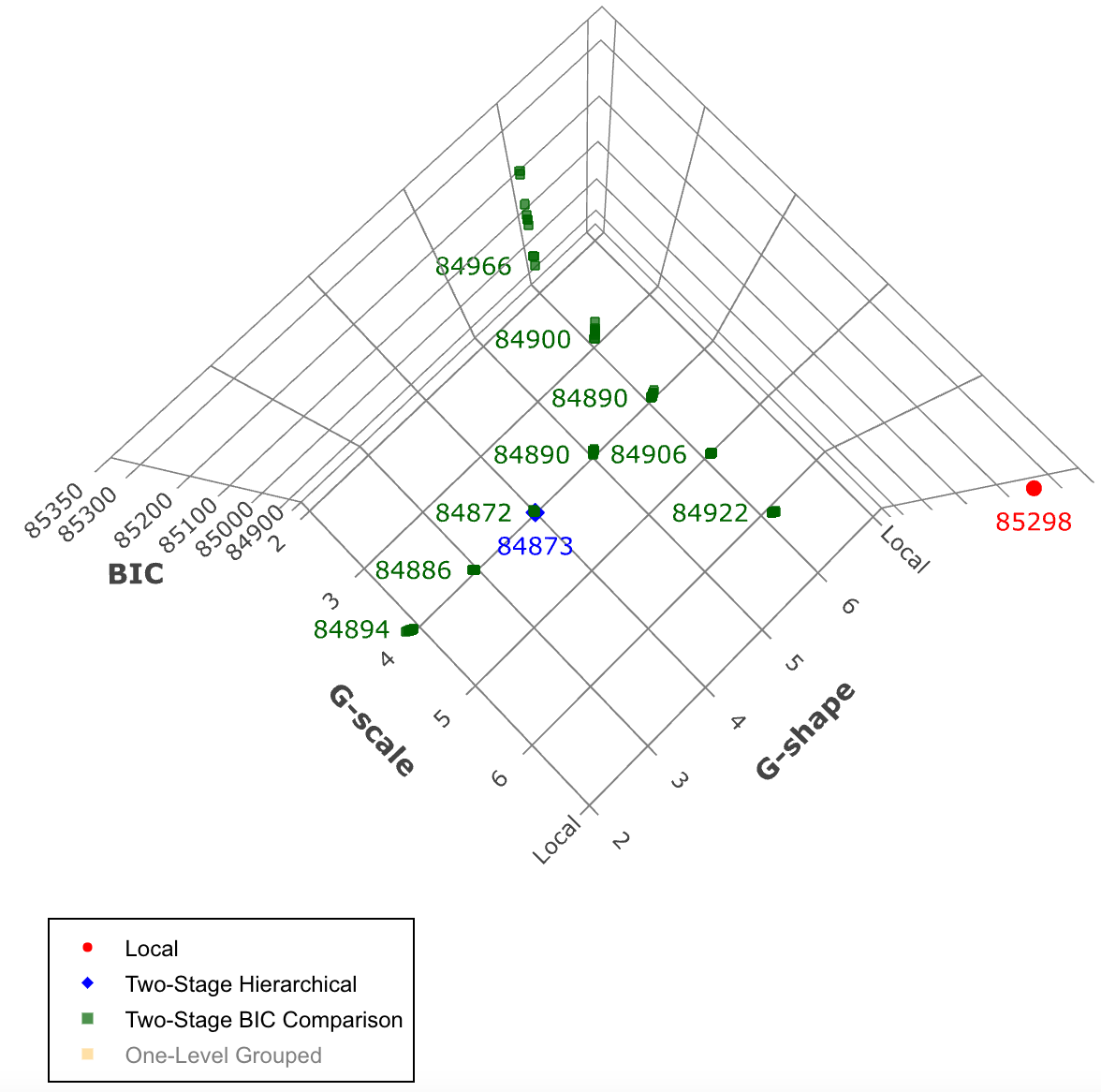}}
    \end{minipage}
    \begin{minipage}[t]{0.49\textwidth}
        \centering        {\includegraphics[width=\linewidth]{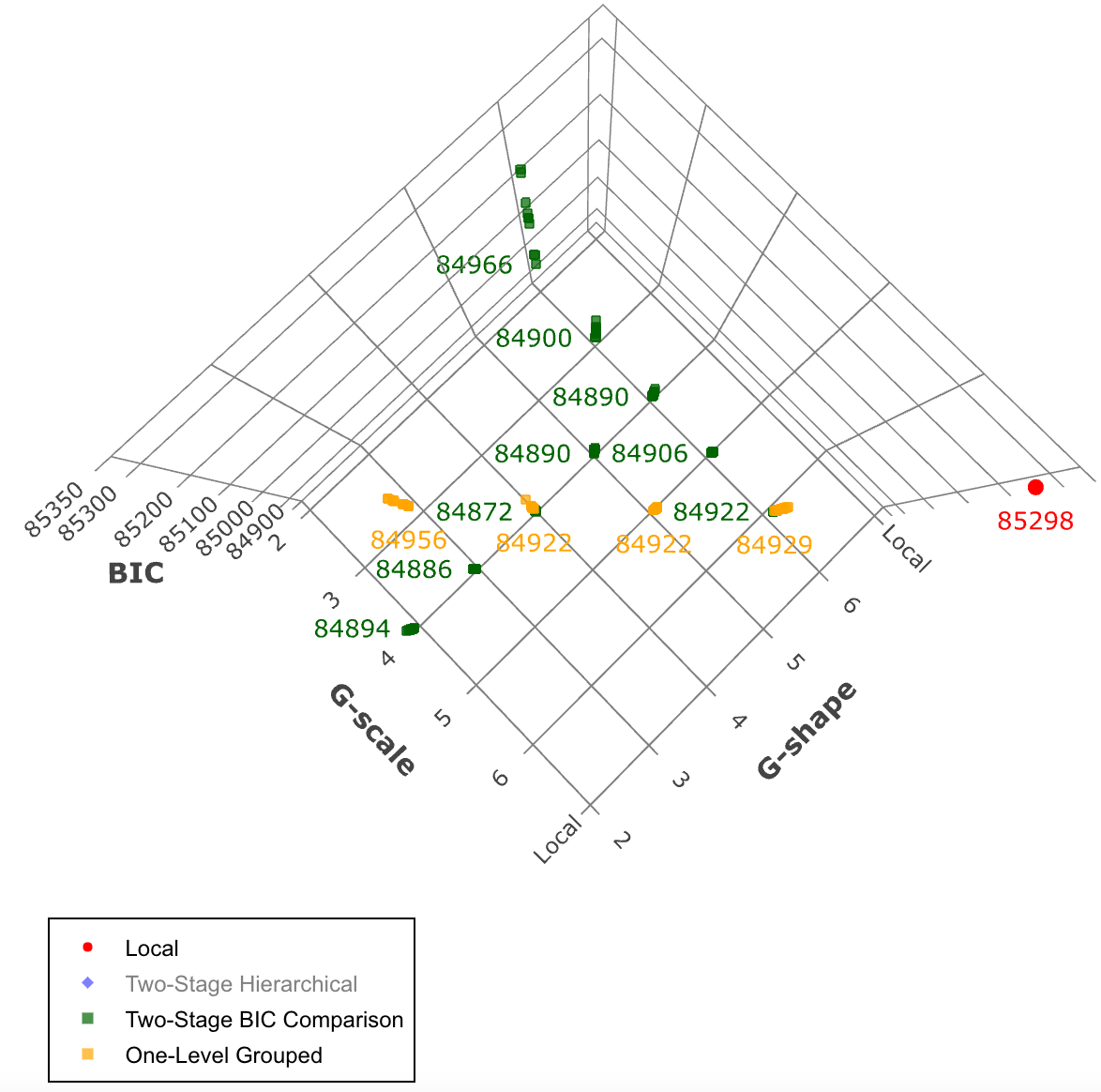}}
    \end{minipage}
    \caption{Comparison of BIC values across a sequence of candidate models. The x-axis and y-axis indicate the values of $\mathcal{G}_{\boldsymbol{\gamma}}$ (\textbf{G-scale}) and $\mathcal{G}_{\boldsymbol{\delta}}$ (\textbf{G-shape}), respectively. With the composite of local models treated as the benchmark, the left panel shows the results under the two-stage approach with BIC comparison and with hierarchical clustering, and the right panel compares the BIC values of the two-level grouped panel GP regression model with the one-level grouped version. For each group dimension pair, we perform 15 runs of the two-step maximization algorithm and 10 runs of the multi-step maximization algorithm. For both of them, only the smallest BIC value obtained is labeled.}
    \label{fig:UDB-BIC-comparison}
\end{figure}

As the two-level grouped model setting already allows additional model flexibility, it is generally advisable not to include too many auxiliary covariates. As an illustrative example, here we incorporate only the average volume of river flow at each station as the covariate to explain the cross-sectional heterogeneity in the scale parameters, where a positive regression coefficient is expected. The shape parameters are assumed to be covariate-independent by convention. We restrict our comparison to the following candidate models: first, we fit a local GP model at each station without incorporating any covariates and then combine them into a composite model, which requires the estimation of 62 parameters; second, we fit the grouped panel GP regression models by running the two-step maximization algorithm multiple times for $\mathcal{G}\in\{3,4,5,6\}$; third, we fit the two-level grouped panel GP regression models over a sequence of group dimension pairs~$(\mathcal{G}_{\boldsymbol{\gamma}},\mathcal{G}_{\boldsymbol{\delta}})$, where we consider both the two-stage approach with BIC comparison and with hierarchical clustering. Specifically, we first set $\mathcal{G}_{\boldsymbol{\delta}}=\lceil\sqrt{31} \rceil=6$ and run the multi-step maximization algorithm multiple times across $\mathcal{G}_{\boldsymbol{\gamma}}\in\{2,3,4,5,6\}$; then, with the $\mathcal{G}^*_{\boldsymbol{\gamma}}$ identified by BIC comparison, we apply the hierarchical clustering approach and run the multi-step maximization algorithm multiple times across each value of $\mathcal{G}_{\boldsymbol{\delta}}\in\{2,3,4,5\}$. A comparison of the BIC's of all identified models is given in the Figure \ref{fig:UDB-BIC-comparison}
. The results are summarized as follows. First, in terms of BIC, all grouped panel GP regression models—regardless of being one-level or two-level grouped—consistently outperform the composite model that combines all local fits. This highlights the overall advantage of our modeling strategy, which integrates regression techniques with flexible data-driven grouping to model peaks over thresholds in panel data. Second, with $\hat{\mathcal{G}}^*_{\boldsymbol{\gamma}} = 4$ fixed, the value of $\mathcal{G}^*_{\boldsymbol{\delta}}$ identified using the hierarchical clustering approach aligns with that obtained by comparing the BIC values from multiple runs of the multi-step maximization algorithm over $\mathcal{G}_{\boldsymbol{\delta}} \in \{2, 3, 4, 5\}$, with both methods suggesting~$\hat{\mathcal{G}}^*_{\boldsymbol{\delta}} = 4$. Moreover, the BIC of the model chosen by hierarchical clustering is very close to that of the optimal model among the 40 (i.e., 10 times at each $\mathcal{G}_{\boldsymbol{\delta}}\in\{2,3,4,5\}$) runs of the multi-step algorithm, where a closer examination reveals that the only difference between the two models lies in the tail group assignment of a single station. These findings confirm the reliability of the hierarchical clustering approach while ensuring the practical convenience of simultaneously estimating $\mathcal{G}^*_{\boldsymbol{\delta}}$ and $\mathbf{A}^*_{\boldsymbol{\delta}}$. Despite its slightly higher BIC, here we choose the model identified by the two-stage approach with hierarchical clustering for further analysis, which is referred to as the chosen model. Finally, as shown in the right panel of Figure~\ref{fig:UDB-BIC-comparison}, the two-level grouped model consistently outperforms the one-level grouped counterpart in terms of BIC, both when comparing the overall best models under each assumption and comparing the best models under the same group dimension pairs (i.e., $\mathcal{G}_{\boldsymbol{\gamma}}=\mathcal{G}_{\boldsymbol{\delta}}=4$ and $\mathcal{G}_{\boldsymbol{\gamma}}=\mathcal{G}_{\boldsymbol{\delta}}=6$). In particular, under $\mathcal{G}_{\boldsymbol{\gamma}}=\mathcal{G}_{\boldsymbol{\delta}}=4$, the optimal group dimension pair identified, the BIC difference of 50 between the best models under each model assumption implies a potentially substantial discrepancy in the latent group structures associated with the $\sigma$-related and $\xi$-related parameters, as explored later.

\begin{figure}[!t]
    \centering    
    \includegraphics[width=14.7cm, height=3.8cm]{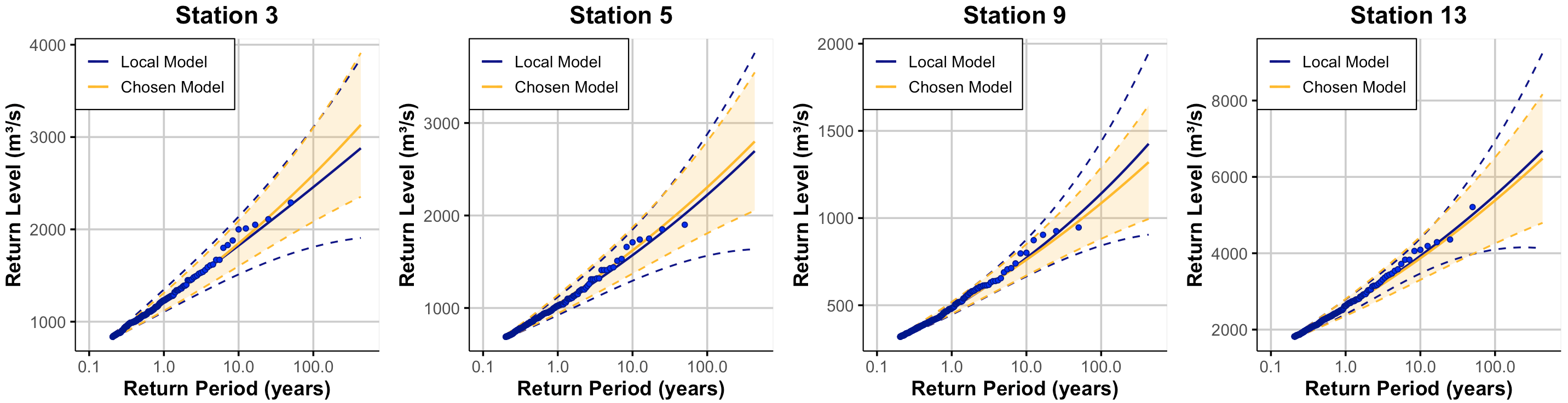} 
    \includegraphics[width=14.7cm, height=3.8cm]{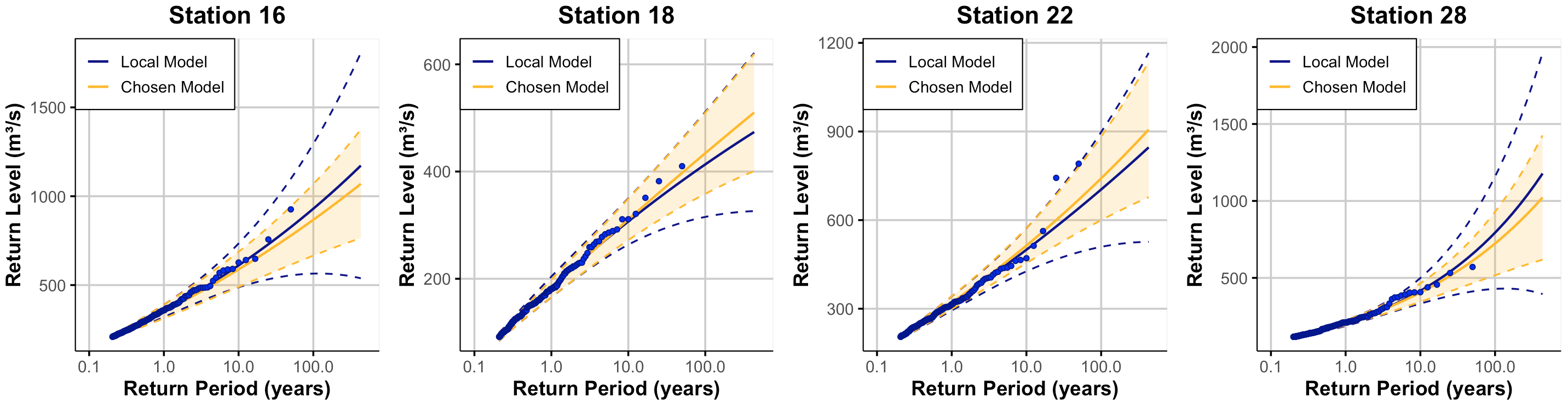} 
    \caption{Return-Level Plots under the group structures identified by the chosen model with estimates from local models used as benchmarks. Dashed lines represent the upper and lower bounds of the 95\% confidence intervals associated with the estimates under each model. The maximum return period included is 500 years.}
    \label{fig:UDB-RL-Plots}
\end{figure}

\begin{figure}[!t]
    \centering    
    \includegraphics[width=15cm, height=6.5cm]{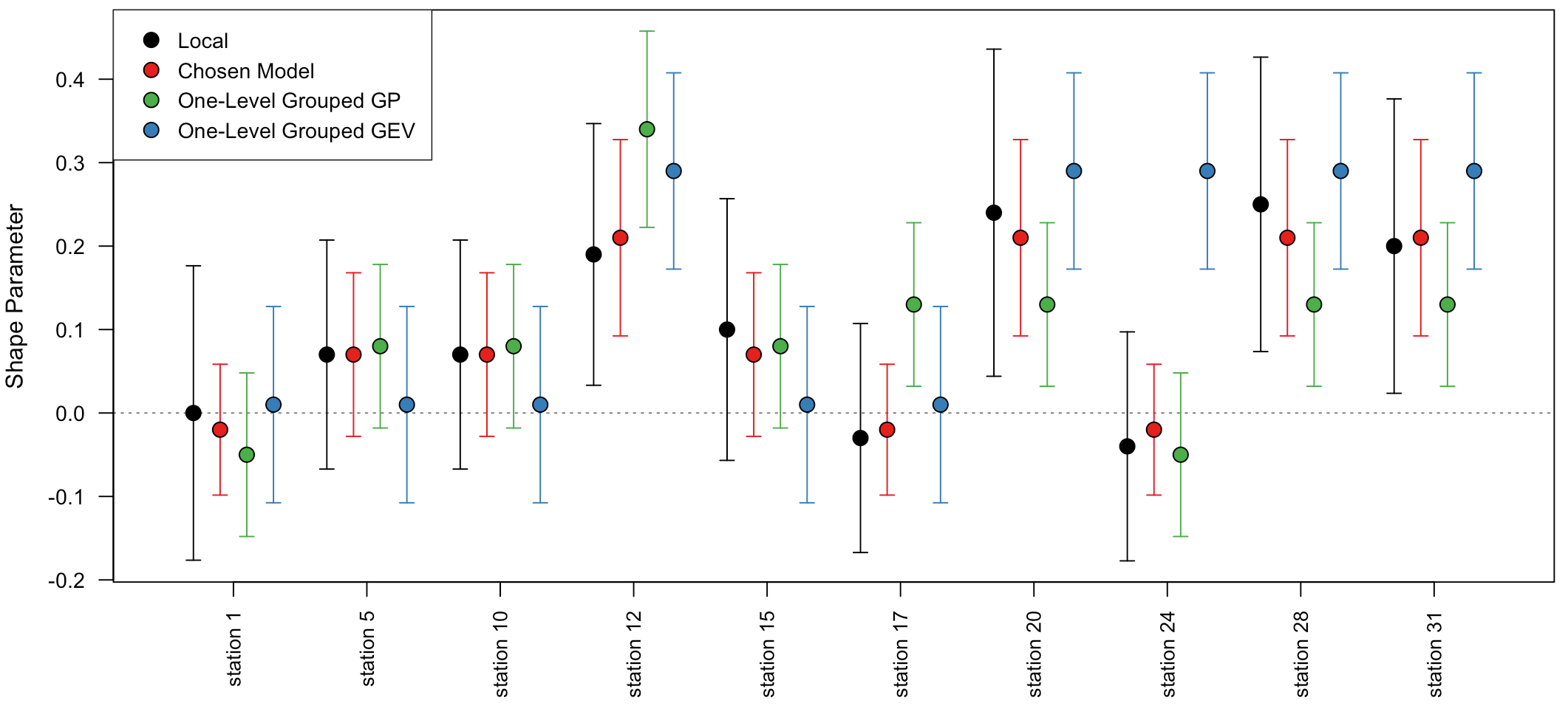}  
    \caption{Comparison of shape parameter estimates across local GP model, the chosen model, the optimal one-level grouped panel GP regression model (in terms of BIC) and the optimal one-level grouped panel GEV regression model across selected stations. The vertical bars indicate the 95\% confidence intervals associated with the estimates.}  
    \label{fig:UDB-shape_est_comp}
\end{figure}

For model diagnostics, we primarily examine the return level plots under the chosen model across the 31 stations, using the return level estimates under the local GP models as benchmarks. A selected subset of the plots is presented in Figure \ref{fig:UDB-RL-Plots} 
 for illustration. We first note that the local GP models achieve good fit at nearly all stations, which validates our choice of thresholds; for a large proportion of these stations, the chosen model yields return level estimates that deviate only slightly or very moderately from those of the local model while substantially reducing the uncertainty associated with the estimates. At certain stations (e.g., stations 3 and 18), the chosen model appears to even better capture the underlying data patterns compared to the local GP model. For the small subset of stations where some lack of fit is observed under the chosen model, various adjustments can be considered. For example, a simple approach is to assign separate group labels for these stations in the $\xi$-related parameter group structure under the chosen model and directly re-estimate the regression parameters. While the updated results under this approach are provided in the supplementary material, we retain the original group structure under the chosen model throughout the rest of our analysis. Next we compare the approximate 95\% confidence intervals for the shape parameters under the chosen model with those under other candidate models. Besides the local models and the best one-level grouped panel GP regression model, we also consider the (one-level) grouped panel GEV regression model introduced in \cite{Dupuis_2023} with the following two modifications. First, we include the area, altitude, and mean slope of the corresponding subcatchment for each station as covariates, all on a log scale as in \cite{Dupuis_2023}, while excluding latitude. When transformed onto a log scale, the latitudes across stations show minimal variation and can easily result in computational singularity in covariance matrix estimation. Second, instead of yearly maxima, we fit the model over the maxima of the three summer months each year. While the two are expected to not differ substantially, the former is more likely to meet the stationarity assumption outlined in the supplementary material of \cite{Dupuis_2023}. The optimal number of groups identified by BIC under this modified model is 2, agreeing with that under the original model used by \cite{Dupuis_2023}. The results of comparison are shown in Figure \ref{fig:UDB-shape_est_comp}
. Restricting attention to the one-level grouped panel GP and GEV regression models, the former tends to offer a mild to moderate additional reduction in the uncertainty of shape parameter estimates relative to the latter, consistent with the motivation for adopting the POT framework. However, compared with the local models, both of them can sometimes suffer from the bias due to the restrictive group structure space they are constrained to explore. In contrast, the chosen model under the two-level grouped panel GP regression model assumption always yields shape estimates that closely align with those from the local models while
significantly reducing the width of the confidence intervals, agreeing with the pattern observed in the return level plots.

\begin{figure}[!t]
    \centering        
    \includegraphics[width=14cm, height=9.7cm]{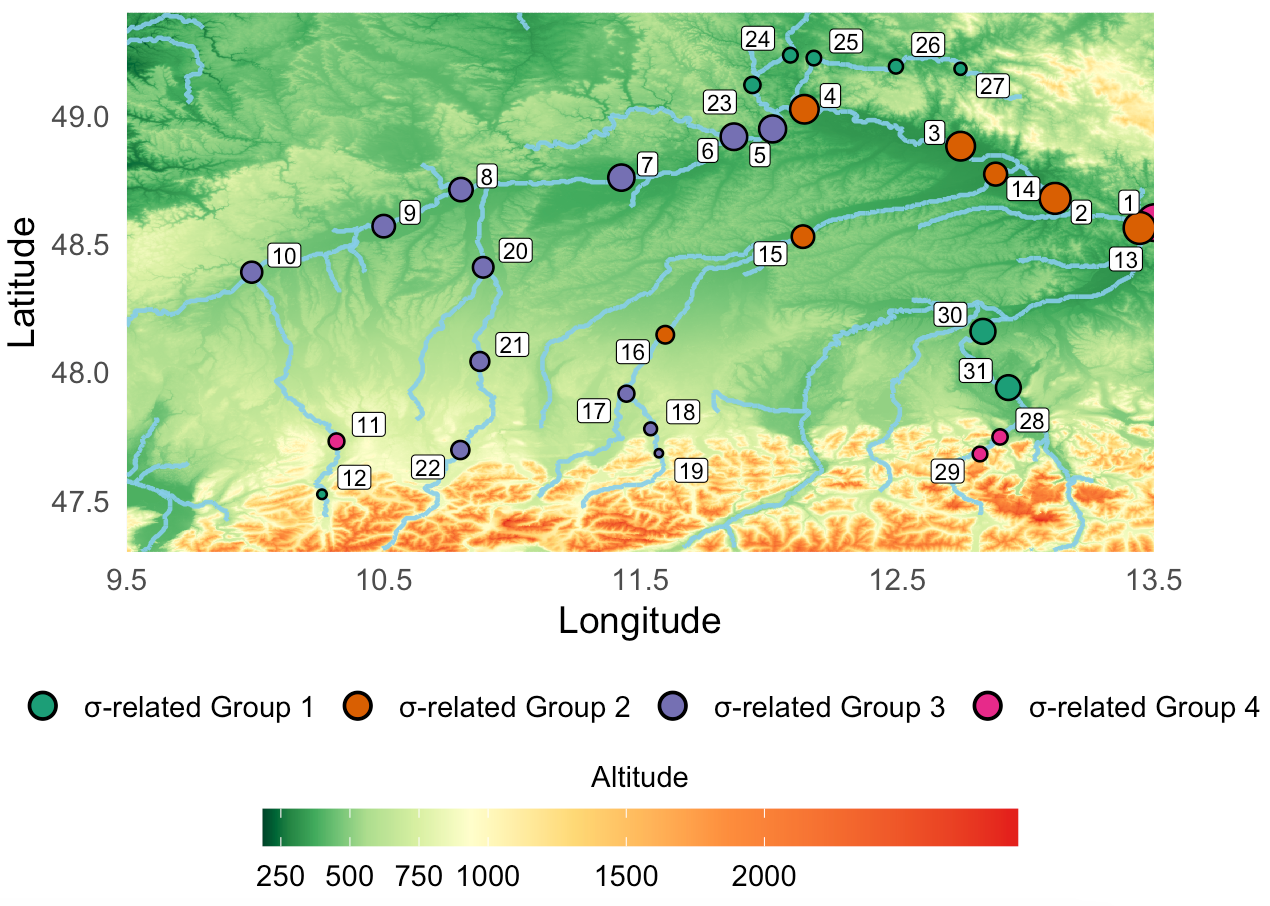}  
    \caption{Identified group structure within the $\sigma$-related regression parameters under the chosen model.}
\label{fig:UDB-Scale-Groups-Chosen}
\end{figure}

\begin{figure}[!t]
    \centering        
    \includegraphics[width=13.2cm, height=7.7cm]{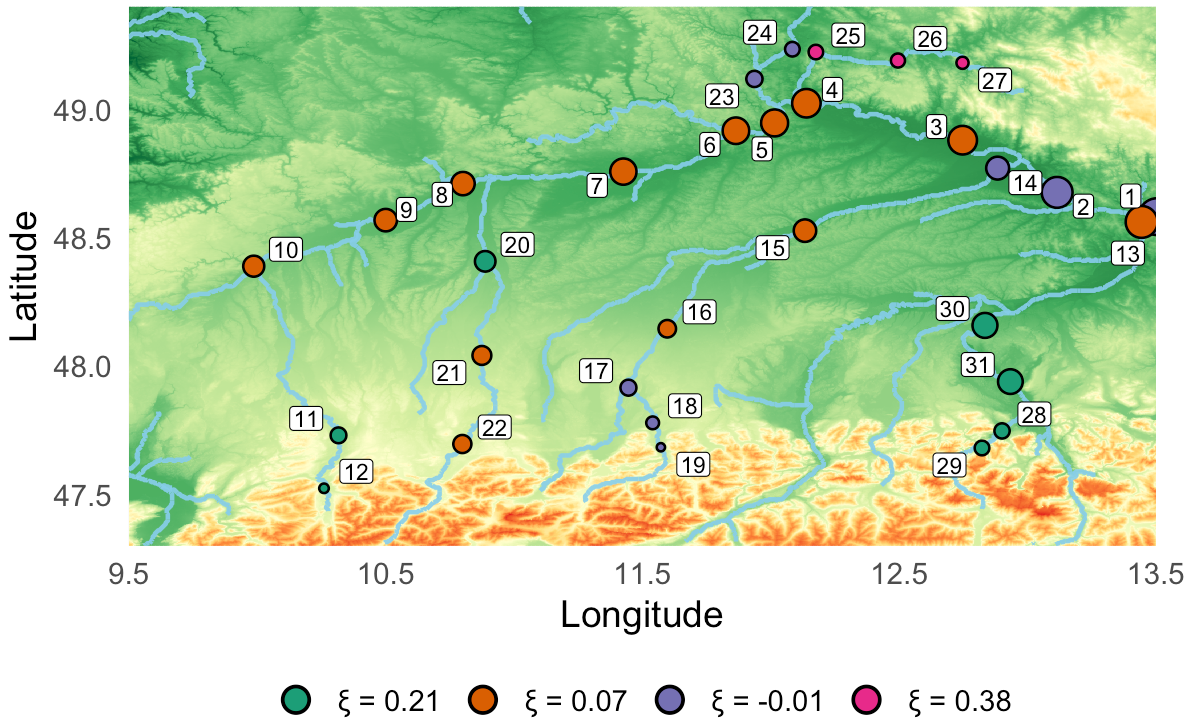} 
    \vspace{0.5cm}
    \includegraphics[width=13.4cm, height=9.5cm]{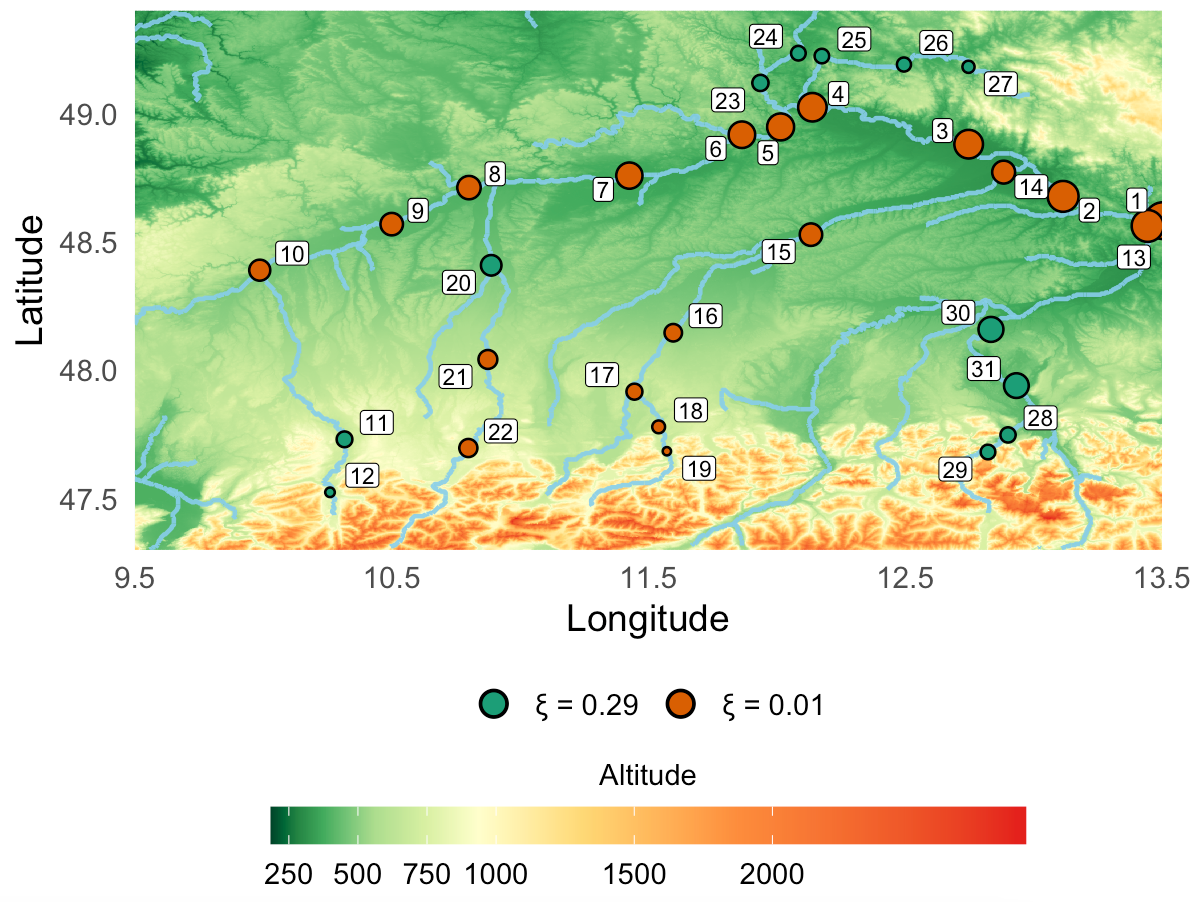} 
    \caption{Identified group structure of the shape parameters under the chosen model (top) and the optimal model identified under the one-level grouped panel GEV regression model (bottom).}
\label{fig:UDB-Shape-Groups}
\end{figure}

We next check the potential interpretations of the identified group structures within the $\sigma$-related and $\xi$-related parameters under the chosen model. For the former, as shown in Figure \ref{fig:UDB-Scale-Groups-Chosen} 
, two large groups of stations emerge in the central region of the upper Danube river basin, which are separated by stations 4 and 5, as well as stations 16 and 17. The five stations in the north of the Danube river are also grouped together. Although the grouping decisions for certain stations in the southwest and southeast of the basin are not straightforward, likely due to the involvement of the covariate(s), the data-driven regionalization produced by our modelling strategy remains overall interpretable in terms of geographical proximity. For the group structure of the shape parameters under the chosen model, to draw a better interpretation, we use that of the optimal one-level grouped panel GEV regression model as a point of reference. As shown in the top panel of Figure \ref{fig:UDB-Shape-Groups} 
, under the chosen model, stations in the central region of the Danube basin still tend to form large clusters, within which a mildly heavy-tailed group is identified. This pattern is consistent with the findings of \cite{asadi2015extremes}, although the estimated tail in our case appears slightly heavier. This discrepancy may be attributed to the inclusion of additional mountain-region stations within that group. As noted in \cite{asadi2015extremes} and also observed here, stations located in mountainous regions tend to exhibit heavier tails, which may be linked to the steep terrain and rapid runoff response in mountain regions. A similar pattern can be observed in the tail group structure under the one-level grouped GEV regression model, as illustrated in the bottom panel of Figure \ref{fig:UDB-Shape-Groups}, which indeed depends on the joint similarity in $\sigma$-related and $\delta$-related parameters. Although a more parsimonious structure—comprising one heavy-tailed group and one light-tailed group—is retained here due to the inclusion of additional covariates, most stations in the heavy-tailed group also belong to the two heavier-tailed groups under the chosen model, and a similar pattern is observed for the light-tailed group. For stations 23 and 24—the only two stations where a discrepancy is observed—a reasonable inference is that their grouping is strongly influenced by the similarity in their $\sigma$-related parameters under the GEV model. The results shown in Figure \ref{fig:UDB-Scale-Groups-Chosen} may serve as a useful reference. We also note that if a prior-knowledge-based regionalization is adopted, as in \cite{asadi2015extremes}, it remains natural to consider stations 23 to 27—all located north of the Danube River—as a single group. This decision will also lead to an estimation result that all of the stations share a similar heavy tail. In this case, our methodology, which classifies stations 23 and 24 into a light-tailed group, demonstrates its ability to detect subtle discrepancies between prior-knowledge-based inference results and the actual patterns revealed by the data.

%=========CONCLUSION=========
\section{Conclusion and future work}
\label{Conclusion}

In summary, we extend the work of \cite{Dupuis_2023} to a two-level grouped panel generalized Pareto regression framework, which is designed to more effectively leverage information from extreme events and to identify models that can achieve greater parsimony and better fit to data at the same time. Clearer interpretations of the identified group structures are also expected. For parameter estimation based on composite likelihood, a dependence-window-based sandwich covariance estimator is formulated, which addresses both the temporal and cross-sectional dependence (conditional on covariates) among univariate threshold excesses. To alleviate the estimation challenges posed by high-dimensional optimization, we propose the use of block coordinate ascent method under our model assumptions. This, in turn,  motivates a multi-step maximization algorithm for jointly estimating the latent group structures and regression parameters. We also recognize the inefficiency of relying on numerous attempts of the multi-step maximization algorithm followed by BIC comparisons for group structure identification. A two-stage approach where the grouping of shape parameters relies on a hierarchical clustering procedure is thus suggested. Through simulation studies, we carefully evaluate the finite-sample performance of the proposed estimation methodology, and satisfactory results are observed. We use an application to the summer discharge data of upper Danube basin to show the advantage of the new modelling framework and the efficacy of the proposed estimation workflow in practice. The two-level grouped model identified using the two-stage approach with hierarchical clustering can achieve a BIC that is very close to that of the optimal model identified using the original BIC comparison approach. Compared with the one-level grouped GP and GEV models, while significantly reducing the standard error of the estimates, it demonstrates a model fit that is either comparable to or better than that of the local GP model across most of the stations. Building on the findings of \cite{asadi2015extremes} and a group panel GEV regression model based on the choices of \cite{Dupuis_2023}, while confirming that the stations located in or near mountainous regions tend to exhibit heavy tails, our methodology is able to identify light-tailed stations within a station cluster that would be estimated to share a common heavy tail under previous methods. This reveals the key advantage of decoupling the group structures for detecting subtle group patterns in parameters.

We discuss potential extensions of our current work from the perspectives of estimation methods and modelling strategies. While retaining the use of MCL estimators, one may consider maximizing a weighted univariate composite likelihood to improve estimation efficiency. In the context of extreme value analysis, it is reasonable to infer that the weight of each excess should depend on the identified clustering structure among them, where clustered excesses should receive lower weights. See \cite{harden2013weighted} for a comprehensive review. We also note that our current covariance estimator incorporates dependence contributions from all pairs of excesses within the pre-specified dependence window as the window ``moves" along time. Sometimes it is reasonable to assume certain degree of sparsity among these contributions, and a penalty-based or a threshold-based approach may apply, thereby reducing the variance associated with standard error estimates. For model and group dimension selection, while the current approach relies on the robustness of BIC against dependence, one may consider using the composite likelihood Bayesian information criterion (clBIC) as an alternative (see \cite{gao2010composite}), which has a better Bayesian interpretation under the composite likelihood framework. However, this choice may lead to computational concerns, as it additionally requires calculating all components of the sandwich estimator for each candidate model. Natural extensions on modelling strategy include adapting our methodology to a three-level grouped model under the block maxima framework and extending the current assumption on group structure to a mixed-effects version, where the effects of covariates vary slightly across subjects within each group but remain centered around a common value. It may also be of interest to investigate a Bayesian version of our model. For a detailed discussion on Bayesian modeling of the GP distribution, see \cite{Bayesian_GP}.

%=========ACKNOWLEDGEMENT=========
\section*{Acknowledgment}
\label{acknowledgement}

The research was funded by the Natural Sciences and Engineering Research Council (NSERC) of Canada.

%=========BIBLIOGRAPHY=========
\newpage
\bibliographystyle{apalike} 
\bibliography{reference} 

\end{document}